\documentclass[12pt,a4paper]{article}
\usepackage{amssymb,amsmath,epsfig}

\newcommand{\eq}{\begin{eqnarray*}}
\newcommand{\qe}{\end{eqnarray*}}
\newcommand{\eqn}{\begin{eqnarray}}
\newcommand{\qen}{\end{eqnarray}}

\newcommand{\ii}{i}

\newcommand{\rrangle}{\rangle\!\rangle}
\newcommand{\bartial}{{\bar{\partial}}}
\newcommand{\Real}{\mathbb{R}}
\newcommand{\Complex}{\mathbb{C}}

\newcommand{\Natural}{\mathbb{N}}
\newcommand{\Integer}{\mathbb{Z}}
\newcommand{\mat}{\begin{pmatrix}}
\newcommand{\tam}{\end{pmatrix}}

\newcommand{\mc}{\mathcal}
\newcommand{\mf}{\mathfrak}

\def\cC{\mc{C}} 
\def\WZNW{{\text{WZNW}}}
\def\WZ{{\text{WZ}}}

\def\inta{{\text{int}}}

\def\Gid{\mc{G}_{\text{id}}}
\def\Allowed{\text{All}}
\def\Rep{\text{Rep}}
\def\Center{\mc{Z}}
\def\id{\text{id}}

\def\tr{\text{tr}}
\def\Tr{\text{Tr}}

\def\embin{\hookrightarrow}
\def\g{\mathfrak{g}}
\def\h{\mathfrak{h}}

\title{\bf Asymmetric Cosets}

\date{}
\author{\bf Thomas Quella$^{a,b}$ and Volker Schomerus$^b$\\[5mm]
  \qquad\small\begin{tabular}{rl}
    $^a$ & Max-Planck-Institut f\"ur Gravitationsphysik\\
         & Albert-Einstein-Institut\\
         & Am M\"uhlenberg 1, D-14476 Golm, Germany\\[3mm]
    $^b$ & Service de Physique Th\'eorique, CEA/DSM/SPhT\\
         & Unit\'e de recherche associ\'ee au CNRS, CEA-Saclay\\
         & F-91191 Gif sur Yvette Cedex, France\\[5mm]
         & \sf quella@aei.mpg.de, vschomer@spht.saclay.cea.fr
  \end{tabular}}

\begin{document}

\maketitle
\vspace{-9.5cm}hep-th/0212119\hfill AEI-2002-091, SPhT-T02/173\vspace{8.5cm}

\begin{abstract}
The aim of this work is to present a general theory of coset 
models $G/H$ in which different left and right actions of $H$ on 
$G$ are gauged. Our main results include a formula for their 
modular invariant partition function, the construction of a 
large set of boundary states and a general description of the 
corresponding brane geometries.
The paper concludes with some explicit applications 
to the base of the conifold and to the time-dependent Nappi-Witten 
background.%
\end{abstract}

\baselineskip17pt

\newpage  
\section{Introduction}

Many interesting models can be obtained as cosets $G/H$ of a compact 
group $G$. Usually, $H$ is identified with a subgroup of $G$ and in 
forming the coset one employs the adjoint action for which $H$ acts 
symmetrically from the left and from the right. Such symmetric 
transformations always possess fixed points (e.g.\ the group unit). 
These  lead to all kinds of singularities of the resulting coset 
geometry, including boundaries and corners.%
\smallskip

It is possible, however, to work with an enlarged class of exactly 
solvable cosets and this is the theme of the following note. The 
idea is to admit different left and right actions of $H$ on $G$. 
Even though conformal invariance imposes strong constraints on 
asymmetric quotients $G/H$, one gains a lot of freedom in model 
building. Some of the interesting new theories possess smooth 
background geometries. One such example is provided by the 
five-dimensional base $SU(2) \times SU(2)/ U(1)$ of the conifold. 
Other models have isolated singularities such as the big-bang 
singularity in the four-dimensional Nappi-Witten geometry $SU(2) 
\times SL(2,\Real)/ \Real \times \Real$.%
\smallskip   

In spite of these interesting features, asymmetric cosets have not 
been studied very systematically in the past. One reason for this is 
that they are typically heterotic, i.e.\ they possess different left  
and right chiral algebras. Among the few publications which deal with 
special cases of asymmetric cosets one may find two early publications 
by Guadagnini et al.\ \cite{Guadagnini:1987ty,Guadagnini:1987qc}. The 
models which are studied in these papers can be applied to the base of 
the conifold as was pointed out some years ago by Pando-Zayas and 
Tseytlin \cite{Pando-Zayas:2000he}. Actions for a wider class of 
asymmetrically gauged WZNW models were written down in \cite{Bars:1992pt}. 
We shall recall below that they are relevant for Nappi-Witten type 
models \cite{Nappi:1992kv}. The latter have been employed recently to 
investigate string theory in time-dependent backgrounds with big-bang 
singularities \cite{Elitzur:2002rt}. Branes in asymmetrically gauged 
WZNW models were also studied in \cite{Walton:2002db} but our analysis
will give boundary theories with a different geometric interpretation.%
\medskip   

The plan of this paper is as follows. In the next section we 
shall present a comprehensive discussion of the bulk theory and, 
in particular, spell out an expression for its modular invariant 
partition function. The third section is then devoted to the 
construction of boundary states for asymmetric cosets. We will 
also identify the subspaces along which the corresponding branes
are localized. All these results on boundary conditions in 
asymmetric coset theories are based on our earlier work 
\cite{Quella:2002ct,Quella:2002ns}. In the final section we 
illustrate the general theory through some important examples. 
These include the base of the conifold and the Nappi-Witten 
type coset background.%
\bigskip 

\noindent {\bf Note added:} While we were preparing this publication, 
G.\ Sarkissian issued a paper that has partial overlap with section
\ref{sc:Examples} below \cite{Sarkissian:2002nq}.%
  
\section{\label{sc:Bulk}The bulk theory}
  
In this first subsection we are going to describe the bulk geometry 
of asymmetric cosets. We will start with a detailed formulation of 
the general setup and of the conditions that conformal invariance 
imposes on the basic data. The origin of the latter can be explained
with the help of the classical actions which we shall briefly recall 
in the second subsection. We then provide expressions for the bulk 
partition functions and establish their modular invariance. Finally, 
we present some examples showing the wide applicability of asymmetric 
cosets. In an appendix to this section we correct some earlier results 
of Guadagnini et al.\ \cite{Guadagnini:1987ty,Guadagnini:1987qc}.%

\subsection{The geometry of asymmetric cosets} 

Two groups $G$ and $H$ enter the construction of a coset $G/H$. Both 
of them are assumed to be reductive so that they split into a product 
of simple groups and $U(1)$ factors. Let the number of these factors 
be $n$ and $r$, respectively, i.e.\ we take $G$ and $H$ to be of the 
form $G=G_1\times\cdots\times G_n$ and $H = H_1 \times \cdots \times 
H_r$. Furthermore, to each factor $G_i$ in the decomposition of $G$ 
we assign a level $k_i$. It is convenient to combine the set of all 
these levels into a vector $k = (k_1,\cdots,k_n)$.%
\smallskip 

Along with the two groups $G$ and $H$ we need to specify an action 
of $H$ on $G$. We take the latter to be of the form $g\mapsto
\epsilon_L(h)g\epsilon_R(h^{-1})$ where $\epsilon_{L/R}:H \rightarrow 
G$ denote two group homomorphisms which descend to embeddings of the 
corresponding Lie algebras. In the usual coset theories $\epsilon_L$ 
and $\epsilon_R$ are the same. An asymmetry in the coset construction 
arises when we drop this condition and allow for two different maps.%
\smallskip 

The coset space $G/H$ consists of orbits under the action of $H$
on $G$, i.e.\ 
\begin{equation*}
  G/H \ = \ G / [\,  g \sim \epsilon_L(h) g \epsilon_R(h^{-1})\, ; \, 
   h \in H\, ]\ \ .
\end{equation*}
To be precise, we should display the dependence on the choice of 
$\epsilon_{L/R}$. But since we consider these maps to be fixed once 
and for all, we decided to suppress them from our symbol $G/H$ for 
the coset space. Let us stress, however, that the geometry is very 
sensitive to the choice of $\epsilon_{L/R}$. We will see this in the 
examples later on.%
\smallskip 

The basic data we have introduced so far, i.e.\ the two groups $G$, 
$H$, the vector $k$ of levels and the maps $\epsilon_L$, $\epsilon_R,$ 
will enter the construction of two-dimensional models with target space 
$G/H$. To ensure conformal invariance, however, these data have to 
obey one important constraint which we can formulate using the notion 
of an ``embedding index'' $x_\epsilon \in $Mat$(n \times r)$ for the
homomorphism $\epsilon: H \rightarrow G$. To define $x_\epsilon$ we 
split $\epsilon$ into a matrix of homomorphisms $\epsilon^{si}:H_s
\embin G_i$ where $s = 1, \dots, r,$ and $i=1, \dots, n,$ run through 
the factors of $H$ and $G$, respectively. The embedding index 
$x_\epsilon  = x = (x^{si})$ is a matrix with elements of the
form\footnote{We use a normalized trace $\Tr=2\,\tr/I$. Here, we denoted
by $\tr$ the matrix trace and by $I$ the Dynkin index of the corresponding
representation. We use the conventions of \cite[pages 58 and 84]{Fuchs:1995}.}  
\begin{equation}
  \label{eq:EmbIndex}
  x^{si}\ =\ \frac{\Tr_i\bigl\{\epsilon^{si}(X)\,\epsilon^{si}(Y)\bigr\}}
              {\Tr_s\bigl\{X\,Y\bigr\}}
  \qquad\text{ for }\qquad X,Y\in\h_s\backslash\{0\}\ \ .
\end{equation}
Observe that the number that is computed by the expression on the right 
hand side does not depend on the choice of the elements $X,Y$. Let us 
also note that the map $\epsilon^{si}$ is allowed to map $H_s$ onto 
the unit element in $G_i$ for some choices of $i$ and $s$. In this 
case, the corresponding matrix element $x^{si}$ vanishes.%
\smallskip

Let us now consider the embedding indices $x_L$ and $x_R$ for the two 
homomorphisms $\epsilon_L$ and $\epsilon_R$. A conformal theory with 
target space $G/H$ exists for our choice of levels $k$, provided that 
the latter obey the following constraint\footnote{If there are two or 
more identical groups, this equation has to hold up to a possible 
relabeling of these groups on one side.}
\begin{equation}
  \label{eq:NoAnomaly}   
  x_L \, k \ = \ x_R \, k \ \ . 
\end{equation}
In other words, the vector of levels must lie in the kernel of $x_L 
- x_R$. For symmetric cosets this condition is trivially satisfied 
with any choice of $k$. Asymmetric cosets, however, constrain the 
admissible levels.%

\subsection{\label{sc:BulkLagrange}The classical action}

Using the basic data we have introduced in the previous subsection we
can write down the classical action of a gauged WZNW model. As usual, 
this consists of several pieces. To begin with, there is the WZNW 
action for the numerator group $G$, 
\begin{equation}
  \label{eq:WZNW}
  \mc{S}_{\WZNW}^{G}\bigl(g|k\bigr)
  \ =\ \sum_{i=1}^n \:\mc{S}_{\WZNW}^{G_i}(g_i|k_i)
\end{equation}
where $g = g_1 \cdot \dots \cdot g_n$. This action is a sum over the 
WZNW actions for the individual groups $G_i$ without any interaction 
terms. These building blocks are given by 
\begin{equation*}
  \mc{S}_{\WZNW}^{G_i}(g_i|k_i)
  \ =\ -\frac{k_i}{4\pi} \int_\Sigma d^2\!z\:\Tr_i\bigl\{\partial g_i g_i^{-1}
   \, \bartial g_i g_i^{-1}\bigr\}  + \mc{S}_{WZ}^{G_i}(g_i|k_i)\ \ .
\end{equation*}
The Wess-Zumino terms are defined as usual in terms of the Wess-Zumino  
three-forms $\omega_i^{\WZ}$. Consistency of the associated quantum 
theories enforces quantization constraints on the levels $k_i$. For 
simply-connected simple constituents $G_i$ the level $k_i$ has to be 
an integer. For the $U(1)$ part and non-simply-connected groups the 
constraints will be different.%
\smallskip

The action functional \eqref{eq:WZNW} is invariant under the ``global''
transformations of the form $g(z,\bar z) \mapsto g_L(z)\, g(z,\bar z) \, 
g_R^{-1}(\bar{z})$ where $g_L(z)$ and $g_R(\bar{z})$ are arbitrary (anti-) 
holomorphic $G$-valued functions. Our subgroup $H$ along with the two 
homomorphisms $\epsilon_{L/R}$ can be used to gauge some part of this 
WZNW symmetry. To this end we consider the model
\begin{equation}
  \label{eq:GaugedWZNW}
  \mc{S}^{G/H}\bigl(g,A,\bar{A}\bigl|\,k,\epsilon_{L/R}\bigr)
  \ =\ \sum_{i=1}^{n}\ \mc{S}_{\WZNW}^{G_i}(g_i|k_i)
       +\sum_{i=1}^n\sum_{s=1}^r\ \mc{S}_{\inta}^{G_i/H_s}
    \bigl(g_i,A_s,\bar{A}_s\bigl|\,k_i,\epsilon_{L/R}^{si}\bigr)\ \ .
\end{equation}
Here, the building blocks of the second term are given by \cite{Bars:1992pt}
\begin{multline}
  \label{eq:Interaction}
  \mc{S}_{\inta}^{G_i/H_s}\bigl(g_i,A_s,\bar{A}_s\bigl|\,k_i,\epsilon_{L/R}^{si}\bigr)
  =\frac{k_i}{4\pi} \int_\Sigma\!d^2\!z\:\Tr_i\bigl\{
       \, 2\,\epsilon_L(\bar{A}_s)\,\partial g_ig_i^{-1} 
          -2\,\epsilon_R(A_s)\, g_i^{-1}\bartial g_i \\[2mm] 
   \qquad\qquad\qquad+2\,\epsilon_L(\bar{A}_s)\, g_i\,\epsilon_R(A_s)\,
         g_i^{-1}  -\epsilon_L(\bar{A}_s)\, \epsilon_L(A_s)
         -\epsilon_R(\bar{A}_s)\,\epsilon_R(A_s)\bigr\}
   \ \ . 
\end{multline}
In this formula we omitted the superscripts $\ ^{si}$ on $\epsilon_{L/R}$. 
The gauge fields $A_s,\bar{A}_s$ take values in the Lie algebra $\mf{h}_s$.
It is not difficult to check that the full action \eqref{eq:GaugedWZNW} is 
invariant under the following set of infinitesimal gauge transformations 
\begin{equation*}
  \begin{split}
   \delta A_s &\ =\ \ii \,\partial\omega_s+\ii \,[\omega_s,A_s]\ \ ,\ \ 
   \delta\bar{A}_s \ = \ \ii \,\bartial\omega_s+\ii \,[\omega_s,\bar{A}_s]\ \ , 
   \\[2mm] 
 \delta g_i &\ =\ \ii \,\epsilon_L(\omega_s)g_i-\ii \,g_i\epsilon_R(\omega_s)\ \ \ \ %
   \text{ for }\ \ \omega_s = \omega_s(z,\bar z) 
  \end{split}
\end{equation*}
provided that the levels $k_i$ obey the constraint \eqref{eq:NoAnomaly}. In fact, 
under gauge transformations the action behaves according to 
\begin{multline*}
  \delta\mc{S}^{G/H}(g|k,\epsilon_{L/R})
  \ =\ \sum_{i=1}^n\sum_{s=1}^r\frac{k_i}{4\pi} \int_\Sigma\!d^2\!z\:\Tr_i
       \bigl\{\, \epsilon_L^{si}(\bar{A}_s)\, \partial\epsilon_L^{si}(\omega_s) 
               -\epsilon_R^{si}(\bar{A}_s)\,\partial\epsilon_R^{si}(\omega_s)\\[2mm] 
               + \epsilon_R^{si}(A_s)\,\bartial\epsilon_R^{si}(\omega_s)
               -\bartial\epsilon_L^{si}(\omega_s)\,\epsilon_L^{si}(A_s)\bigr\}
\end{multline*}
and so it vanishes whenever eq.\ \eqref{eq:NoAnomaly} holds true. We have therefore 
shown that the data introduced above indeed label different two-dimensional conformal 
field theories.%

\subsection{\label{sc:BulkAlgebra} Exact solution: modular invariant partition function}

Our aim now is to present a few elements of the exact solution. We shall 
begin with some remarks on the relevant chiral algebras and then address 
the construction of the modular invariant partition function for our 
asymmetric coset theories.%
\smallskip 

In the following let us denote the chiral algebra of the WZNW model for the 
group $G$ and levels $k_i$ by $\mc{A}(G)$. This algebra is generated by 
a sum of affine Lie algebras with levels $k_i$, one for each factor in the 
decomposition of the reductive group $G$. The two maps $\epsilon_{L/R}$ 
give rise to two embeddings of the chiral algebra $\mc{A}(H)$ into 
$\mc{A}(G)$. Let us note that $\mc{A}(H)$ is generated by a 
sum of affine algebras, one for each factor in the product $H = H_1 \times
\cdots \times H_r$. The levels of these affine algebras form a vector
$(k^\prime_s)_{s=1,\dots,r}$ whose entries are related to the levels of $\mc{A}(G)$
by $k^\prime= x_{L/R}\,k$ (matrix notation). Our assumption \eqref{eq:NoAnomaly} means that 
$\epsilon_{L/R}$ give rise to two (possibly different) embeddings of the {\em same} 
chiral algebra $\mc{A}(H)$ into $\mc{A}(G)$. Given these embeddings, 
we employ the usual GKO construction to obtain two coset algebras $\mc{A} = 
\mc{A}(G/H,\epsilon_L)$ and $\mc{\bar{A}} = \mc{A}(G/H,\epsilon_R)$
which form the left and right chiral algebras of the asymmetric 
coset model. Note that these two chiral algebras can be different if the two 
maps $\epsilon_L$ and $\epsilon_R$ are not the same. In this sense, asymmetric 
coset models of the kind that we consider in this note are heterotic conformal 
field theories.%
\smallskip  
  
The state space of any conformal field theory decomposes into representations of 
the chiral algebra. Our task here is to find a combination of these representations 
which does not only reflect the geometry of the target space $G/H$ but is at the 
same time also consistent from a conformal field theory point of view. The second 
requirement means that the vacuum must be unique and that the partition function 
is modular invariant.%
\smallskip
  
The first condition, namely the relation of our exact solution to the space $G/H$, 
implies that in the limit of large levels $k$ the space of ground states has to 
reproduce the space of functions on $G/H$. Actually, we can turn this around for 
a moment and use the harmonic analysis of $G/H$ to get some ideas about the 
structure of the state space. To this end, let us recall that the algebra 
$\mc{F}(G)$ of functions on $G$ may be considered as a $G\times G$-module 
under left and right regular action. The Peter-Weyl theorem states that this 
module decomposes into irreducibles according to
\begin{equation*}
  \mc{F}(G) \ =\ \bigoplus\: V_\mu\otimes V_{\mu^+}\ \ ,
\end{equation*}
where $\mu^+$ is the conjugate of $\mu$. Since we want to divide $G$ by the 
action of $H$ it is convenient to decompose the space of function on $G$ into 
representations of $H$. The space of function on $G/H$ is then obtained as the 
$H$-invariant part of $\mc{F}(G)$. We easily find
\begin{equation}
  \label{eq:GeomPartition}
  \begin{split}
  \mc{F}(G)
  &\ \cong\ \bigoplus\: V_\mu\otimes V_{\mu^+}\\[2mm]  
  &\ \cong\ \bigoplus {(b_L)_\mu}^a\,{(b_R)_{\mu^+}}^{c^+}
     \ V_a\otimes V_{c^+}\\[2mm]
  &\ \cong\ \bigoplus {(b_L)_\mu}^a\,{(b_R)_{\mu^+}}^{c^+}\,{N_{ac^+}}^d\ V_d
\ \ .
  \end{split}
\end{equation}
The symbols $b_{L/R}$ denote the branching coefficients of the inclusion
$\epsilon_{L/R}(H) \embin G$. The tensor product coefficients ${N_{ac^+}}^d$ 
for the decomposition of the tensor product of representations of $H$ enter 
when we restrict the action of $H\times H$ to its diagonal subgroup $H = H_D$.
Taking the invariant part of \eqref{eq:GeomPartition} corresponds to putting
$d=0$ or, equivalently, $a=c$ and hence we have shown that  
\begin{equation*}
  \mc{F}(G/H) \ = \ \text{Inv}_{H_D} \bigl(\, \mc{F}(G)\, \bigr) \ \cong \
    \bigoplus \: {(b_L)_\mu}^a\,{(b_R)_{\mu^+}}^{a^+} \ \ .
\end{equation*}
This is the space that we want to reproduce from the ground states of our 
exact solution when we send the levels to infinity. With a bit of experience 
in coset chiral algebras and their representation theory it is not too
difficult to come up with a good proposal for the conformal field theory 
state space that meets this requirement.%
\smallskip 

The rough idea is to replace the branching coefficients ${b_\mu}^a$ by 
coset sectors $\mc{H}_{(\mu,a)}^{G/H}$. But this rule is a bit too simple
and has to be refined in several directions. To build in all the additional 
subtleties, we need a bit of preparation. For simplicity we consider the
sector of the left moving chiral algebra only.%
\smallskip

In the following we label sectors of $\mc{A}(G)$ by $\mu,\nu, \dots,$ and we use 
the letters $a,b, \dots,$ for sectors of $\mc{A}(H)$. Let us recall that the two 
sets of sectors admit an action of the group centers $\mc{Z}(G)$ and $\mc{Z}(H)$, 
respectively. This action may be diagonalized by the corresponding modular 
S-matrices,  
\begin{align*}
  S_{J\mu\,\nu}^G&\ =\ e^{2\pi\ii \, Q_J(\nu)}\:S_{\mu\nu}^G&
  \text{for }&J\in\mc{Z}(G)\ \ 
\end{align*}
where $Q_J(\nu)=h_J+h_\mu-h_{J\mu}$ are the so-called monodromy charges. An analogous 
statement holds for the action of the center $\mc{Z}(H)$. In a coset sector $(\mu,a)$ 
the labels $\mu,a$ form an entity and as such they have to transform identically under 
the common center
\begin{equation*}
  \Gid(L)\ =\ \bigl\{(J,J^\prime)\in\mc{Z}(G)\times\mc{Z}(H)\,\bigl|\,J=
  \epsilon_L(J^\prime)\bigr\}\ \ .
\end{equation*}
Not all the labels $(\mu,a)$ fulfill this requirement. What remains is the set
\begin{equation*}
  \Allowed(G/H)_L\ =\ \bigl\{\,(\mu,a)\,\bigl|\,Q_J(\mu)=Q_{J^\prime}(a)
  \text{ for all }(J,J^\prime)\in\Gid(L)\,\bigr\}
\end{equation*}
of allowed coset labels. It turns out that elements in the set $\Allowed(G/H)_L$ 
which are related by the action of $\Gid$ correspond to the same coset sector. 
The set of sectors for the coset chiral algebra is therefore given by  
$\Rep(G/H)_L=\Allowed(G/H)_L\bigr/\Gid(L)$. This observation motivates the term 
``field identification group'' for the common center $\Gid(L)$. The same 
constructions can be performed for the right chiral algebra. But note that 
in general the resulting expressions will not coincide.%
\smallskip

Having introduced all these notions from the representation theory of 
coset chiral algebras we are finally able to spell out our proposal for 
the state space, 
\begin{equation}
  \label{eq:GeneralHilbert}
  \mc{H}^{G/H}\ =\ %
  \bigoplus_{[\mu,a]\in\Rep(G/H)}
  \mc{H}_{(\mu,a)}^{(G/H)_L}\otimes\mc{\bar{H}}_{(\mu,a)^+}^{(G/H)_R}\ \ .
\end{equation}
where the set $\Rep(G/H)$ is defined by 
\begin{equation}
  \label{eq:RepGH}  
  \begin{split}  
  \Rep(G/H) &\ =\ \Allowed(G/H)\Bigl/\Gid\ \ \text{ with } \\[2mm]
  \Allowed(G/H) &\ =\ \Allowed(G/H)_L \cap \Allowed(G/H)_R   
 \ \ , \ \ \Gid\ =\ \Gid(L)\cap\Gid(R) \ \ .  
  \end{split}
\end{equation} 
Note that the field identification group $\Gid$ admits a natural interpretation
as the stabilizer of the action $g\mapsto\epsilon_L(h)g\epsilon_R(h)^{-1}$, 
i.e.\  
\begin{equation*}
  \Gid \ = \ \bigl\{\bigl(J,J^\prime\bigr)\,\bigl|\,
    J^\prime\in\Center(H),\ J=\epsilon_L(J^\prime)=\epsilon_R(J^\prime)\in\Center(G)\bigr\}\ \ .
\end{equation*}
In writing our formula \eqref{eq:GeneralHilbert} we implicitly assumed 
that the action of the field identification group $\Gid$ on $\Allowed(G/H)$ 
possesses no fixed points, i.e.\ that all orbits $[\mu,a]$ have the same length. 
It should be stressed that fixed points for the action of $\Gid(L/R)$ on 
$\Allowed(G/H)_{L/R}$ are not ruled out by this assumption.%
\medskip

As we mentioned before, our proposal \eqref{eq:GeneralHilbert} for the state 
space has to pass a number of tests before we can accept it as a candidate 
for the state space of our conformal field theory. From our discussion above
it is not difficult to see that at large level, the space of ground states 
coincides with the space of functions on $G/H$. Moreover, taking the quotient 
with respect to $\Gid$ in eq.\ \eqref{eq:RepGH} ensures that there is a unique
vacuum in $\mc{H}^{G/H}$. Hence, it only remains to demonstrate that our 
Ansatz also leads to a modular invariant partition function. To this end     
it is convenient to write the partition function in the form 
\begin{equation*}
  Z(q,\bar{q})\ =\ \frac{1}{|\Gid|}\sum_{\mu,a}\text{P}_{L}^{G/H}(\mu,a)\,\text{P}_{R}^{G/H}(\mu^+,a^+)
     \:\chi_{(\mu,a)}^{(G/H)_L}(q)\bar{\chi}_{(\mu,a)^+}^{(G/H)_R}(\bar{q})\ \ .
\end{equation*}
The factor $1/|\Gid|$ in front of this expression removes a common factor from 
the whole expression in such a way that the vacuum characters possess a trivial 
prefactor. The summation in the previous expression runs over all labels $\mu$ 
and $a$ and we enforce the restriction to the allowed coset labels by inserting 
the projectors
\begin{equation*}
  \text{P}_{L/R}^{G/H}(\mu,a)\ =\ \frac{1}{|\Gid(L/R)|}\sum_{(J,J^\prime)\in\Gid(L/R)}\:
   e^{2\pi\ii(Q_J(\mu)-Q_{J^\prime}(a))}\ \ .
\end{equation*}
\smallskip 

It is now rather straightforward to compute how this partition function behaves
under the modular transformation $S$ that replaces $q = \exp(2\pi i \tau)$ by 
$\tilde{q} = \exp(-2\pi i /\tau)$, 
\begin{equation*}
  SZ(q,\bar{q})\ =\sum_{\mu,a,\nu,\lambda,b,c} 
   \frac{\text{P}_{L}^{G/H}(\mu,a)\,\text{P}_{R}^{G/H}(\mu^+,a^+)}{|\Gid|}
    \ S_{\mu\nu}^G\bar{S}_{ab}^H\bar{S}_{\mu^+\lambda^+}^GS_{a^+c^+}^H\ %
     \chi_{(\nu,b)}^{(G/H)_L}(q)\bar{\chi}_{(\lambda,c)^+}^{(G/H)_R}(\bar{q})\ \ .
\end{equation*}
We would like to use unitarity of the S-matrices to simplify this expression. 
But before we can do so, we have to get rid of the projectors. To this end we   
insert the explicit formulas for the projectors in terms of monodromy charges 
and then pull the latter into the S-matrices by shifting their indices with the 
action of simple currents. This gives
\begin{equation*}
  \begin{split}
  SZ(q,\bar{q})
  &\ =\ \frac{1}{|\Gid|\cdot|\Gid(L)|\cdot|\Gid(R)|}
     \sum_{(J_1,J_1^\prime),(J_2,J_2^\prime)}
     \sum_{\mu,a,\nu,\lambda,b,c}\\
  &\hspace{2cm} S_{\mu J_1\nu}^G\bar{S}_{aJ_1^\prime b}^H
     \bar{S}_{\mu^+ J_2\lambda^+}^GS_{a^+J_2^\prime c^+}^H\ %
     \chi_{(\nu,b)}^{(G/H)_L}(q)\bar{\chi}_{(\lambda,c)^+}^{(G/H)_R}(\bar{q})\ \ .
  \end{split}
\end{equation*}
Now we are able to perform the sum over $\mu$ and $a$ to obtain
\begin{equation*}
  \begin{split}
  SZ(q,\bar{q})
  &\ =\ \frac{1}{|\Gid|\cdot|\Gid(L)|\cdot|\Gid(R)|}
     \sum_{(J_1,J_1^\prime),(J_2,J_2^\prime)}\sum_{\nu,\lambda,b,c}
  \delta_{J_1\nu}^{J_2^{-1}\lambda}\delta_{J_2^{\prime-1} c}^{J_1^\prime b}\ %
     \chi_{(\nu,b)}^{(G/H)_L}(q)\bar{\chi}_{(\lambda,c)^+}^{(G/H)_R}(\bar{q})\ \ .
  \end{split}
\end{equation*}
At this stage we may resum the label. Then we see that part of the prefactor
cancels and we are left with
\begin{equation*}
  SZ(q,\bar{q})
  \ =\ \frac{1}{|\Gid|}\sum_{\nu,b}
       \chi_{(\nu,b)}^{(G/H)_L}(q)\bar{\chi}_{(\nu,b)^+}^{(G/H)_R}(\bar{q})\ \ .
\end{equation*}
This is exactly the behavior modular invariance requires from our partition 
function. Note that the restriction to allowed coset labels is implicitly 
contained in the previous expression since coset characters vanish if the 
relevant branching selection rule is not satisfied. It is obvious that our
partition function is also invariant under modular $T$-transformations which
send $\tau$ to $\tau+1$.%
\smallskip
  
\subsection{Special cases and examples}

Our general construction includes a number of interesting special cases. 
The most familiar examples are the Nappi-Witten background \cite{Nappi:1992kv} 
and the $T^{pq}$-spaces \cite{Pando-Zayas:2000he}. Both of them belong to a 
distinguished class of asymmetric cosets for which we introduce the notion 
of ``generalized automorphism type''. In the last subsection we will briefly 
discuss one example of an asymmetric coset that is not of this type.%

\subsubsection{Asymmetric cosets from automorphisms}

The simplest setup for asymmetric cosets that one can imagine is one in 
which the left and right embeddings are related by automorphisms. More 
precisely, we are thinking of situations in which the left homomorphism
$\epsilon_L= \epsilon$ is related to $\epsilon_R= \Omega_G^{-1}\circ 
\epsilon\circ\Omega_H$ by composition with two automorphisms $\Omega_G$ 
and $\Omega_H$ of $G$ and $H$, respectively. Let us notice that the concatenation 
of an embedding with an automorphism gives another embedding with the same 
embedding index.\footnote{In our terminology, automorphisms do not only have  
to respect the group multiplication but also the Killing form (or an other 
invariant form in terms of which the model is defined).} This observation 
guarantees the validity of the anomaly cancellation condition 
\eqref{eq:NoAnomaly}.%
\smallskip

For the explicit construction of the state space \eqref{eq:GeneralHilbert}
we have to know the centers $\Gid(L)$ and $\Gid(R)$ in detail. Note that 
every element $\bigl(\epsilon(J^\prime),J^\prime\bigr)\in\Gid(L)$ is mapped 
to an element $\bigl(\Omega_G^{-1}\circ\epsilon(J^\prime),\Omega_H^{-1}(J^\prime)\bigr)
=\bigl(\Omega_G^{-1}\circ\epsilon\circ\Omega_H(\Omega_H^{-1}(J^\prime)),\Omega_H^{-1}
(J^\prime)\bigr)\in\Gid(R)$ by the action of the pair $(\Omega_G^{-1},\Omega_H^{-1})$.
The right center is thus the image of the left center, $\Gid(R)=(\Omega_G^{-1},
\Omega_H^{-1})\bigl(\Gid(L)\bigr)$, and the common center is the intersection of 
these two sets. Similarly the allowed coset labels are related by $\Allowed
(G/H)_R=(\Omega_G^{-1},\Omega_H^{-1})\bigl(\Allowed(G/H)_L\bigr)$. To prove this 
statement one employs the invariance property $Q_{\Omega_G(J)}\bigl(\Omega_G(\mu)
\bigr)=Q_J(\mu)$ of the monodromy charges and the analogous statement for the 
subgroup $H$.%
\smallskip

These observations enable us to find a rather explicit expression for the state 
space. In our example the general formula \eqref{eq:GeneralHilbert} can be 
simplified due to the fact that left and right moving chiral algebra are isomorphic.
We will therefore express the state space in terms of quantities of the left 
chiral algebra. All we need to do is to replace the coset representations
$\mc{H}^{(G/H)_R}_{(\mu,a)}$ through $\mc{H}^{(G/H)_L}_{(\Omega_G(\mu),
\Omega_H(a))}$. By construction, the latter is non-trivial if and only if the 
first one was. We can also express the action of the common center on these 
labels. If we combine these facts we finally arrive at
\begin{equation*}
  \mc{H}^{G/H}\ =\ \bigoplus_{[\mu,a]\in\Rep(G/H)}\mc{H}_{(\mu,a)}^{(G/H)_L}
  \otimes\mc{\bar{H}}_{(\Omega_G(\mu),\Omega_H(a))^+}^{(G/H)_L}\ \ .
\end{equation*}
Let us emphasize once more that the coset sectors are {\em both} defined with
respect to the {\em same} embedding $\epsilon$ in this expression. The asymmetry
enters in the explicit appearance of the twists of labels and in an (implicit)
reduction of labels over which we sum.%
\smallskip

The most prominent example of asymmetric cosets of the type considered in this
subsection is provided by the Nappi-Witten background \cite{Nappi:1992kv}.
It is obtained as a coset of the product group $G=SL(2,\Real)\times SU(2)$
with respect to some abelian subgroup $H=\Real\times\Real$. In this case, the 
automorphism $\Omega_G$ is trivial while $\Omega_H$ exchanges the two factors 
of $\Real$. The model will be discussed in detail in section \ref{sc:Examples}.%

\subsubsection{\label{sc:ExamplesTPQ}Examples of GMM-type}

Let us now consider a slightly more complicated family of examples in which the 
numerator group is a product $G_1\times G_2$ of two groups $G_1$ and $G_2$ which 
possess a common subgroup $H$. Our aim is to describe the coset $G_1\times G_2/H$
where the first homomorphism $\epsilon_L = e \times \epsilon_2$ embeds $H$ into 
the group $G_2$ and $\epsilon_R = \epsilon_1 \times e$ sends elements of $H$ into 
$G_1$. The Lagrangian description of such models was developed by Guadagnini, 
Martellini and Mintchev (GMM) more than fifteen years ago \cite{Guadagnini:1987ty,
Guadagnini:1987qc}. In appendix \ref{ap:GMM} we show how their results can be 
recovered from the more general expression \eqref{eq:GaugedWZNW}. We also use the 
opportunity to correct some statements of GMM concerning the current algebra relations
and the validity of the affine Sugawara / coset construction for this type of coset
models.%
\smallskip

The Lagrangian treatment of appendix \ref{ap:GMM} and algebraic intuition
lets us suspect that the coset model is manifestly heterotic with chiral
algebras given by
\begin{equation*}
  \mc{A}\Bigl((G_1)_{k_1}\otimes\bigl((G_2)_{k_2}/H_{k}\bigr)\Bigr)
  \otimes
  \overline{\mc{A}\Bigl(\bigl((G_1)_{k_1}/H_{k}\bigr)\otimes(G_2)_{k_2}\Bigr)}\ \ .
\end{equation*}
One can easily see that the field identification group for the coset $G_1\times G_2/H$ 
is given by 
\begin{equation*}
  \Gid\ =\ %
  \bigl\{(0,0,J^\prime)\,\bigl|\,(0,J^\prime)\in\Gid(G_1/H)\cap\Gid(G_2/H)\bigr\}\ \ .
\end{equation*}
  The allowed coset labels consist of triples $(\mu,\alpha,a)$ such that
  $(\mu,a)$ and $(\alpha,a)$ are allowed for the $G_1/H$ and $G_2/H$ cosets,
  respectively. Coset representations are then obtained by dividing out the
  field identifications $\Gid$. The resulting state space simply reads
\begin{equation*}
  \mc{H}\ =\ \bigoplus_{[\mu,\alpha,a]\in\Rep(G/H)}\ %
         \mc{H}_\mu^{G_1}\otimes\mc{H}_{(\alpha,a)}^{G_2/H}
         \otimes
         \bar{\mc{H}}_{(\mu,a)^+}^{G_1/H}\otimes\bar{\mc{H}}_{\alpha^+}^{G_2}\ \ .
\end{equation*}
It reflects the fact that in both the left and the right moving algebra one still
finds a residual current symmetry.%
\smallskip

For the physical applications we are particularly interested in a special
choice of product group and subgroup, $G_1=G_2=SU(2)$ and $H=U(1)$. Under these
circumstances the GMM-model describes five-dimensional non-Einstein $T^{pq}$
spaces \cite{Pando-Zayas:2000he}. The special case $p=q=1$ admits a direct
interpretation as the base of the conifold (see, e.g., \cite{Klebanov:1998hh}).
This example will be discussed in detail in section \ref{sc:Examples}.%

\subsubsection{Asymmetric cosets of non-automorphism type}
  
In the last two subsections we discussed examples of asymmetric cosets which are of 
rather special form. Recall that for the first case, the left and right embeddings
were simply related by automorphisms. An interesting generalization of this setup 
involves choosing a chain of subgroups $H = U_1 \subset U_2 \subset \dots \subset 
U_N = G$ along with left and right embeddings which are pairwise related through 
automorphisms. If an asymmetric coset falls into this wider class, we will say that
it is of ``generalized automorphism type''. Note that the GMM coset models belong to 
this family. To see how this works, let us introduce a subgroup $U_2 = H\times H$ 
which sits in between $H$ and $G=G_1\times G_2$. Given such an intermediate group
we first embed $H$ into either the second or first factor of $H\times H$ and then 
continue by embedding $H\times H$ into $G$. In this scenario, the left and right 
embeddings from $H$ to $H \times H$ are related by the permutation automorphism 
of $H \times H$ and the left and right embedding from $H \times H$ to $G$ are 
even identical.%
\smallskip

Asymmetric coset models of generalized automorphism type are heterotic with respect
to their maximal chiral algebras, i.e.\ the algebra of holomorphic chiral fields is
not isomorphic to the algebra of anti-holomorphic fields (unless the model is of 
automorphism type). On the other hand, their chiral algebras possess a smaller 
common chiral subalgebra for which the whole theory is still rational. This property 
will enable us in the next section to write down a large number of boundary states
for asymmetric cosets of generalized automorphism type.%
\smallskip 

Before we proceed to the discussion of boundary conditions, however, we would like 
to provide at least one example of an asymmetric coset that is not of generalized 
automorphism type. To this end we consider once again the product $G=G_1\times G_2$ 
of two simple Lie groups $G_i$ which possess a common subgroup $H$. One can 
define an action of $H$ on $G$ which is based on the embeddings of the following 
special form $\epsilon_L(h)=\bigl(\epsilon_1(h),\epsilon_2(h)\bigr)$ and 
$\epsilon_R(h)=\bigl(\epsilon_1^\prime(h),\id\bigr)$. The corresponding matrices 
of embedding indices are denoted by $(x_1,x_2)$ and $(x_1^\prime,0)$. If we can 
now find levels such that the condition  $k=x_1k_1+x_2k_2=x_1^\prime k_1$ is 
obeyed, then there exists an associated asymmetric coset model with chiral algebra
\begin{equation*}
  \mc{A}\Bigl(\bigl((G_1)_{k_1}\times(G_2)_{k_2}\bigr)/H_{k}\Bigr)
  \otimes
  \overline{\mc{A}\Bigl(\bigl((G_1)_{k_1}/H_{k}\bigr)\times(G_2)_{k_2}\Bigr)}\ \ .
\end{equation*}
For models of this type we were not able to find a common chiral subalgebra for 
which the theory stays rational. An very explicit example is obtained using the 
inclusion $\mf{su}(2)_{4k^\prime}\embin\mf{su}(3)_{k\prime}\oplus
\mf{su}(3)_{3k^\prime}$. In fact, there are two embeddings of $\mf{su}(2)$ into 
$\mf{su}(3)$ at our disposal with embedding indices $1$ and $4$, respectively. 
If we employ the embedding with index $1$ for $\epsilon_i$ and choose 
$\epsilon_1^\prime$ such that it has embedding index $4$, then the anomaly 
cancellation condition is satisfied.%

\section{\label{sc:Boundary}The boundary theory}
  
In this section we will construct boundary states for the asymmetric coset 
theories. The heterotic nature of these models will force us to break part 
of the bulk symmetry. But for a large class of asymmetric cosets we will be
able to identify smaller chiral symmetries for which the boundary theory 
remains rational. Our discussion starts with a short reminder on maximally 
symmetric and symmetry breaking branes on group manifolds. We will then 
argue that some of the symmetry breaking branes on $G$ can descend to the 
asymmetric coset and we will identify the localization of these branes in 
$G/H$. Formulas for the boundary states and the partition functions of the 
boundary theories will be provided at the end of the section.%
  
\subsection{Branes on group manifolds} 

Among the branes on group manifolds, maximally symmetric branes are 
distinguished since they preserve the whole chiral current algebra 
symmetry. The construction of maximally symmetric boundary conditions 
in the WZNW model requires to choose some gluing automorphism $\Omega$ 
of the chiral algebra $\mc{A}(G)$ so that we can glue holomorphic and 
anti-holomorphic currents along the boundary. Before we describe a few
results from boundary conformal field theory of the corresponding branes, 
let us briefly look at the geometric scenario these boundary conditions 
are associated with. It is by now well known that branes constructed 
with $\Omega= \id$ are localized along conjugacy classes \cite{Alekseev:1998mc}. 
The general case has an equally simple and elegant interpretation 
\cite{Felder:1999ka}. Note that gluing automorphisms $\Omega$ for the current 
algebra $\mc{A}(G)$ are associated with automorphisms of the finite 
dimensional Lie algebra $\mf{g}$ which, after exponentiation, give rise 
to an automorphism $\Omega^G$ of the group $G$. One can then show that 
maximally symmetric branes are localized along the following {\em 
twisted conjugacy classes} in the group manifold, 
\begin{equation*}
  \cC^\Omega_u \ := \ \bigl\{\,  g \, u\,  \Omega^G(g^{-1})\ \bigl|\ g \in G 
  \, \bigr\} \ \ .
\end{equation*}
The subsets $\cC^\Omega_u \subset G$ are parametrized through equivalence 
classes of group elements $u$ where the equivalence relation between two 
elements $u,v \in G$ is given by: $u \sim_\Omega v$ iff $v \in C^\Omega_u$. 
One should think of $u$ as a coordinate that describes the transverse 
position of the brane on the group manifold. In the exact conformal field
theory, these coordinates are quantized.%
\smallskip

The algebraic description of maximally symmetric D-branes was developed
in \cite{Birke:1999ik} (see also \cite{Behrend:1999bn}). Their boundary
states are labeled by representations of the twisted Kac-Moody algebra
which may be constructed from the Lie algebra $\mf{g}$ using the
automorphism $\Omega$. They are specific linear combinations of
certain generalized coherent (or Ishibashi) states,
\begin{equation*}
  |u\rangle\ =\ \sum_{\Omega(\mu)=\mu}\:\frac{{\psi_u}^\mu}{\sqrt{S_{0\mu}}}\:|\mu\rrangle\ \ . 
\end{equation*}
As usual, the generalized coherent states only implement the gluing 
conditions for the currents and there is one such state for each 
$\Omega$-symmetric combination of irreducible $\mf{\hat{g}}$-representations 
in the charge conjugate state space of the WZNW theory. The coefficients 
${\psi_u}^\mu$ in the previous formula are directly related to the one-point 
functions of bulk fields in the boundary theories and explicit expressions 
can be found in the literature \cite{Birke:1999ik}. From the boundary states
one can compute the partition function 
\begin{equation*}
  Z_{uv}\ =\ \sum_{\nu\in\Rep(G)}\:{\bigl(n_\nu\bigr)_v}^u\:\chi_\nu
        \ =\ \sum_{\nu\in\Rep(G)}\sum_{\mu=\Omega(\mu)}\:
 \frac{{\bar{\psi_u}}^\mu\,{\psi_v}^\mu\,S_{\nu\mu}}{S_{0\mu}}\:\chi_\nu
\end{equation*}
for each pair of labels $u,v$. The numbers ${\bigl(n_\nu\bigr)_v}^u\in\Natural_0$ 
are the twisted fusion rules of $\mf{\hat{g}}$. For details of the construction
we refer the reader to the existing literature.%
\medskip

In addition to these maximally symmetric branes, a large class of 
symmetry breaking branes has been obtained in \cite{Quella:2002ct}.  
Their geometry was identified later in \cite{Quella:2002ns}. The
construction of these branes requires to choose a chain of groups 
$U_s, s = 1, \dots, N,$ along with homomorphisms $\epsilon_s: U_s 
\rightarrow U_{s+1}$ (we set $U_N = G$). The latter are again assumed 
to induce embeddings of the corresponding Lie algebras. Furthermore, 
one has to select an automorphism $\Omega_s$ on each group $U_s$. 
Given these data, it is possible to construct a set of branes  which 
preserve an $U_1$ group symmetry. These are localized along the 
following sets   
\begin{equation} 
  \label{prodC}
  \begin{split}  
  \cC^{\underline \Omega}_{\underline \epsilon; \underline u}
  &\ =\ \cC^N_{u_N} \, \cdot\,  \cC^{N-1}_{u_{N-1}}\,  \cdot \, \dots\,  
     \cdot\,  \cC^1_{u_1} \ \subset \ G \ \ \ \text{ where } \\[2mm]
  \cC^s_{u_s}&\ =\ \Omega_N \circ \epsilon_{N-1} \circ \dots 
     \circ \Omega_{s+1} \circ  
     \epsilon_s ( \cC^{\Omega_s}_{u_s} ) \ \subset \ 
     G \ \ \ \text{ for } \ \ 
     u_s \ \in \ U_s
  \end{split}
\end{equation}
and $\cC^N_{u_N} = \cC^{\Omega_N}_{u_N}$ for $u_N \in G$. The 
$\cdot$ indicates that we consider the set of all points in $G$ 
which can be written as products (with group multiplication) of 
elements from the various subsets. One should stress that branes
may be folded onto the subsets \eqref{prodC} such that a given
point is covered several times. This phenomenon has been observed
for a special case in \cite{Maldacena:2001ky}. For maximally
symmetric branes on ordinary adjoint cosets the previous geometry reduces
to simpler expressions which have been found before
\cite{Gawedzki:2001ye,Elitzur:2001qd}.%
\smallskip 

To illustrate this abstract construction, we show how to recover the 
symmetry breaking branes on $SU(2)$ that were found in \cite{Maldacena:2001ky}. 
In this case, we choose a chain of length $N=2$ and set $U_1 = U(1)$.
Let us then fix the automorphism $\Omega_1$ on $U(1)$ to be the inversion 
$\Omega_1 (\eta) = \eta^{-1}$ for all $\eta \in U(1)$. The automorphism 
$\Omega_2$ of $SU(2)$ is assumed to be trivial and $\epsilon_1$ can be 
any embedding of $U(1)$ into $SU(2)$. With these choices, the twisted 
conjugacy classes $\cC^{\Omega_1}$ fill the whole one-dimensional circle 
$U_1$. When we multiply points of the circle with elements in the spherical 
conjugacy classes $\cC^{\Omega_2} = \cC^{\id}$ of $SU(2)$ the resulting set 
sweeps out a three-dimensional subspace of $SU(2)$ which can degenerate to a 
1-dimensional circle. Hence, for this very special example, we reproduce 
exactly the findings of \cite{Maldacena:2001ky}.%

\subsection{Branes in asymmetric cosets}

Having constructed maximally symmetric and symmetry breaking branes in 
the group $G$, our strategy now is to investigate which of these branes
can pass down to the asymmetric coset $G/H$. Geometrically, this is not 
too hard to understand. In fact, the natural idea is to look at all the 
symmetry breaking branes which are obtained from chains starting with 
$U_1 = H$ and end at $U_N = G$ and to impose an extra condition on the 
choice of the automorphisms $\Omega_s$ and the homomorphisms $\epsilon_s$ 
so as to reflect the action of $H$ on $G$ in the coset construction. 
Explicitly, the conditions on $\Omega_s$ and $\epsilon_s$ read    
\begin{equation}
  \label{cond1} 
  \epsilon_L\ =\ \epsilon_{N-1} \circ\cdots\circ\epsilon_1\ \ ,
\end{equation}
and 
\begin{equation}
  \label{cond2}
  \epsilon_R \ =\ \Omega^{U_N}\circ\epsilon_{N-1}\circ\Omega^{U_{N-1}}
  \circ\cdots\circ\Omega^{U_2}\circ\epsilon_1 \circ\Omega^{U_1}\ \ ,
\end{equation}
Our claim is that the subsets \eqref{prodC} that are obtained from chains 
$(U_s,\Omega_s)$ with homomorphisms $\epsilon_s$ pass down to subsets on 
the asymmetric coset $G/H$, provided that the data of the chain are related 
to the data $\epsilon_{L/R}$ of the asymmetric coset $G/H$ by eqs. \eqref{cond1}
and \eqref{cond2}. We believe that the branes that are obtained in this way 
are the only ones that possess a rational boundary theory.%

\subsection{Boundary states and partition function}

We now turn to the exact solution of the boundary conformal field theories 
which are used to describe the branes we talked about in the previous subsection. 
Our assumptions on the existence of a chain of embeddings and its properties 
guarantee that the resulting theories are rational with respect to a chiral symmetry 
\begin{equation*}
  \mc{A} \ =\ %
  \mc{A}(U_N/U_{N-1})\otimes\cdots\otimes\mc{A}(U_3/U_2)\otimes\mc{A}(U_2/U_1)\ \ .
\end{equation*}
Let us stress that the left and right chiral algebra are isomorphic after 
symmetry reduction while this did not have to be the case before.
For the rest of this section, let us restrict to embedding chains of length 
$N=3$. This does not only cover all the examples to be discussed later on, but 
it also simplifies our notations. The extension to the general case will be 
straightforward. We will also set $\Omega_1 = \id = \Omega_3$ and write $U_2 
= U$.%
\smallskip

To begin with, it is convenient to rewrite the bulk partition function of the 
asymmetric coset model in terms of characters for the chiral algebra $\mc{A}$ 
that is left unbroken by the boundary condition. With our simplifying assumptions, 
this partition function becomes   
\begin{equation}
  \label{eq:ModularInvariant}
  \mc{H}\ =\ %
  \bigoplus_{[\mu,a]\in\Rep(G/H)}\ \bigoplus_{\alpha,\beta\in\Rep(U)}
            \mc{H}_{(\mu,\alpha)}^{G/U}\otimes
            \mc{H}_{(\alpha,a)}^{U/H}\otimes
            \mc{\bar{H}}_{(\mu,\beta)^+}^{G/U}\otimes
            \mc{\bar{H}}_{(\Omega(\beta),a)^+}^{U/H}\ \ .
\end{equation}
We had to include the automorphisms $\Omega$ in one of the coset representations
because in the original formulation of the symmetry reduction left and right
chiral algebra are just isomorphic, not identical. By explicit insertion of
$\Omega$ we are able to formulate the theory in terms of one single chiral 
algebra $\mc{A}$.%
\smallskip

To construct boundary states we have to find the symmetric part of
the Hilbert space \eqref{eq:ModularInvariant}. From the $G/U$ cosets
we obtain the condition $\alpha\equiv\beta$ modulo field identification
of the form $(e,J)\in\Gid(G/U)$. From the $U/H$ cosets one arrives at
$\alpha=\Omega(\beta)$. This is due to the fact that elements of the field
identification group $\Gid(U/H)$ can not have the form $(J,e)$. The first
condition then translates into $\alpha=J\Omega(\alpha)$. We will assume that
this condition can only be fulfilled for $\Omega(\alpha)=\alpha$.%
\footnote{This condition can be non-trivial only if there exist elements 
in the center of $H$ which are mapped to the unit element by both 
$\epsilon_L$ and $\epsilon_R$.}  Generalized coherent  states  
$|\mu,\alpha,a\rangle\!\rangle$ for this setup are labeled by 
triples $\mu,\alpha,a$ such that 
\begin{eqnarray*}
 (\mu,\alpha)\ \in \ \Allowed(G/U) \ \ \ & , &  \ \ \ 
 (\alpha,a) \ \in \ \Allowed(U/H)\ \ , \\[2mm] 
 \Omega(\alpha)  & = &  \alpha \ \ .
\end{eqnarray*}
In addition we have to identify these generalized coherent states according 
to the identification rule 
\begin{equation*}
  |J\mu,\alpha,J^\prime a\rangle\!\rangle\sim|\mu,\alpha,a\rangle\!\rangle
  \ \ \text{for}\ \ (J,J^\prime)\in\Gid\ \ .
\end{equation*}
 
Let ${\psi_z}^\alpha$ be the structure constants of twisted D-branes in the 
target space $U$. Whenever the tupel $(\rho,z,r)$ satisfies the selection rule
$Q_J(\rho)=Q_{J^\prime}(r)$ for all elements $(J,J^\prime)\in\Gid$
we then may define boundary states for the asymmetric coset by
\begin{equation*}
  |\rho,z,r\rangle\ =\ %
  \sum \text{P}(\mu,\alpha)\text{P}(\alpha,a)\ \frac{S_{\rho\mu}^G}{\sqrt{S_{0\mu}^G}}
  \ \frac{{\psi_z}^\alpha}{S_{0\alpha}^U}
  \ \frac{\bar{S}_{ra}^H}{\sqrt{S_{0a}^H}}
  \ |\mu,\alpha,a\rangle\!\rangle\ \ .
\end{equation*}
We note that this is a consistent prescription since the formula does not depend 
on the specific representative of the Ishibashi states. Also, we may implement 
the identification of boundary states
\begin{equation*}
  |J\rho,z,J^\prime r\rangle\ \sim\ |\rho,z,r\rangle
  \ \ \text{for}\ \ (J,J^\prime)\in\Gid\ \ .
\end{equation*}
Using world-sheet duality it is not difficult to derive formulas for the 
boundary partition functions. As usual we start from the following expression 
involving the coefficients of boundary states 
\begin{equation*}
  Z\ =\ \sum \text{P}(\mu,\alpha)\,\text{P}(\alpha,a)\ \frac{\bar{S}_{\rho_1\mu}^GS_{\rho_2\mu}^G}{S_{0\mu}^G}
  \ \frac{{\bar{\psi}_{z_1}}^\alpha{\psi_{z_2}}^\alpha}{S_{0\alpha}^US_{0\alpha}^U}\ 
  \frac{S_{r_1a}^H\bar{S}_{r_2a}^H}{S_{0a}^H}\ \chi_{(\mu,\alpha)}^{G/U}
  \chi_{(\alpha,a)}^{U/H}(\tilde{q})\ \ 
\end{equation*}
and  perform the modular S transformation to obtain
\begin{equation*}
  Z\ =\ \sum \text{P}(\mu,\alpha)\,\text{P}(\alpha,a)\ \frac{\bar{S}_{\rho_1\mu}^GS_{\rho_2\mu}^GS_{\nu\mu}^G}{S_{0\mu}^G}
  \ \frac{{\bar{\psi}_{z_1}}^\alpha{\psi_{z_2}}^\alpha\bar{S}_{\beta\alpha}^US_{\gamma\alpha}^U}{S_{0\alpha}^US_{0\alpha}^U}\ \frac{S_{r_1a}^H\bar{S}_{r_2a}^H\bar{S}_{ba}^H}{S_{0a}^H}\ \chi_{(\nu,\beta)}^{G/U}\chi_{(\gamma,b)}^{U/H}(q)\ \ .
\end{equation*}
We now want to pass to an unrestricted sum over $\mu,\alpha,a$ ($\alpha$ still has 
to be symmetric). This can be achieved if we express the projectors in terms monodromy 
charges and pull the corresponding simple currents into the S matrices. We are thus lead to
\begin{equation*}
  Z\ =\ \frac{1}{|\Gid^{G/U}|\ |\Gid^{U/H}|}\sum \frac{\bar{S}_{\rho_1\mu}^GS_{\rho_2\mu}^GS_{J_1\nu\mu}^G}{S_{0\mu}^G}
  \ \frac{{\bar{\psi}_{z_1}}^\alpha{\psi_{z_2}}^\alpha\bar{S}_{J_1^\prime\beta\alpha}^U
  S_{J_2\gamma\alpha}^U}{S_{0\alpha}^US_{0\alpha}^U}\ \frac{S_{r_1a}^H\bar{S}_{r_2a}^H
  \bar{S}_{J_2^\prime ba}^H}{S_{0a}^H}\ \chi_{(\nu,\beta)}^{G/U}\chi_{(\gamma,b)}^{U/H}(q)\ \ .
\end{equation*}
The expression may be evaluated directly by means of the Verlinde formula. The final result is
\begin{equation*}
  \begin{split}
  Z&\ =\ \frac{1}{|\Gid^{G/U}|\ |\Gid^{U/H}|}\sum\ N_{\rho_1^+,\rho_2,J_1\nu}
      N_{(J_1^\prime\beta)^+,J_2\gamma}^{\delta}{\bigl(n_\delta\bigr)_{z_1}}^{z_2}
      N_{r_1r_2^+}^{J_2^\prime b}
      \ \chi_{(\nu,\beta)}^{G/U}\chi_{(\gamma,b)}^{U/H}(q)\\
   &\ =\ \sum\ N_{\rho_1^+,\rho_2,\nu}
      N_{\beta^+\gamma}^{\delta}{\bigl(n_\delta\bigr)_{z_1}}^{z_2}
      N_{r_1r_2^+}^{b}
      \ \chi_{(\nu,\beta)}^{G/U}\chi_{(\gamma,b)}^{U/H}(q)\ \ .
  \end{split}
\end{equation*}
Let us remark that this spectrum is consistent with the proposed geometric interpretation. 
The relevant computations are left to the reader (see \cite{Quella:2002ns} for a closely 
related analysis).%

\section{\label{sc:Examples}Examples}

To illustrate the abstract formulas we presented in this work we will now
study three important examples. Our discussion starts with a short analysis 
of D-branes in the parafermionic cosets $SU(2)/U(1)$ and branes therein. We 
then proceed to the spaces $T^{pq}$ generalizing the base of the conifold.
The section concludes with a detailed investigation of branes in the 
Nappi-Witten background.%

\subsection{\label{sc:Parafermions}Branes in the parafermion background}

Parafermion theories arise from cosets of the form $SU(2)/U(1)$. There 
exist two choices of how the $U(1)$ subgroup can be gauged: adjoint 
(vectorial) and axial gauging. D-branes for these models have been 
worked out in \cite{Maldacena:2001ky,Walton:2002db} and we will not 
have anything new to see in this example. Our purpose is merely to 
introduce some of the tools that help to illustrate the bulk and 
brane geometries in specific examples. Recovering the geometry of the 
so-called $A$ and $B$ branes in parafermion theories from our general 
theory will be rather easy.%
\smallskip

\begin{figure}
  \centerline{\input{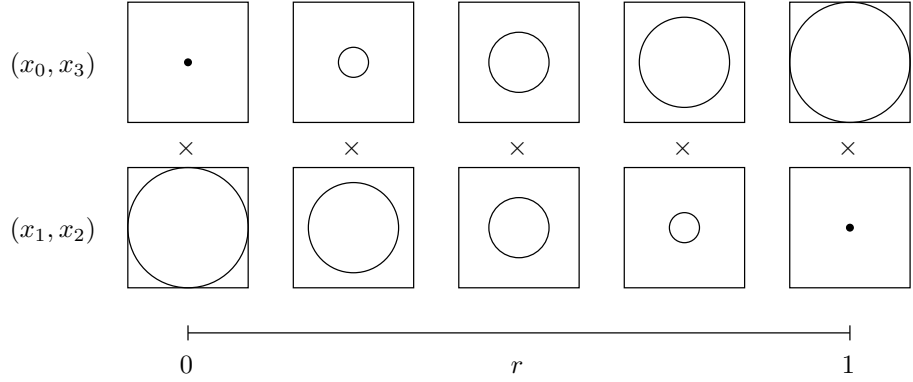}}
  \caption{\label{fig:SU2}The group manifold $SU(2)$ as a fibre over the unit intervall.}
\end{figure}

\begin{figure}
  \centerline{\input{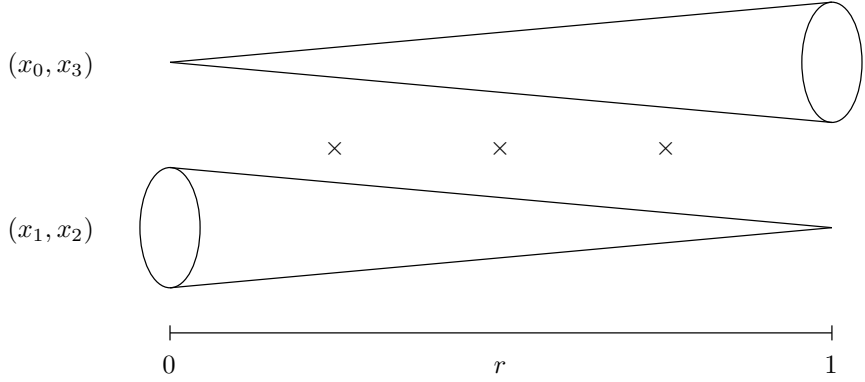}}
  \caption{\label{fig:SU2alt}A second illustration of the group manifold $SU(2)$.}
\end{figure}

The first ingredient in our discussion is the $SU(2)$ group manifold 
itself. It will be useful for us to parametrize it in terms of two 
complex coordinates $z_1,z_2$, 
\begin{equation*}
  g\ =\ \mat z_1&z_2\\-\bar{z}_2&\bar{z}_1\tam\qquad\text{ with }\ 
  z_1,z_2\in\Complex \ \text{ and }\ |z_1|^2+|z_2|^2=1\ \ .
\end{equation*}
To picture this space, we define the quantity $r=|z_1|$ which takes 
values in the interval $0\leq r\leq 1$. Over each point $r \in [0,1]$  
the group manifold fibers into the direct product $S^1_r\times 
S_{\sqrt{1-r^2}}^1$ of two circles with radii $r$ and $\sqrt{1-r^2}$, 
respectively. Hence, we find 
\begin{equation*}
  SU(2)\ =\ \bigl\{(z_1,z_2)\,\bigl|\,|z_1|^2+|z_2|^2=1\bigr\}
  \ =\ %
  \bigl\{\bigl(r\,e^{i\phi_1},\sqrt{1-r^2}\,e^{i\phi_2}\bigr)
  \bigl|0\leq r\leq1\bigr\}\subset\Complex^2\ \ .
\end{equation*}
If we identify the complex numbers with euclidean $2$-planes according to
$z_1=x_0+ix_3$ and $z_2=x_1+ix_2$ we arrive at the figures \ref{fig:SU2}
and \ref{fig:SU2alt}.%
\smallskip

Next we turn to the $U(1)$ subgroup and its embeddings into $SU(2)$. For the 
left embedding $\epsilon_L$ we shall use the map  
\begin{equation*}
  \epsilon_L:\ \ e^{i\tau}\ \mapsto\ \bigl(e^{i\tau},0\bigr)\ \ .
\end{equation*}
The vector and axial gaugings arise from two different choices of the right 
homomorphism $\epsilon_L$ which we choose to be 
\begin{equation*}
  \epsilon^{v/a}_R:\ \ e^{i\tau}\ \mapsto\ \bigl(e^{\pm i\tau},0\bigr)\ \ .
\end{equation*}
The associated actions of $U(1)$ on $SU(2)$ assume a rather simple form in the 
coordinates $(z_1,z_2)$, 
\begin{align*}
  \bigl(z_1,z_2\bigr)&\mapsto\bigl(z_1,\,e^{2i\tau}z_2\bigr)
  \quad\text{(vector)}&\text{and}&&
  \bigl(z_1,z_2\bigr)&\mapsto\bigl(e^{2i\tau}z_1,z_2\bigr)
  \quad\text{(axial)}\ \ .
\end{align*}
In descending to the coset geometry, it is convenient to fix the gauge such 
that either $z_1$ or $z_2$ is a positive real number. We thus arrive at the 
expressions
\begin{equation*}
  \begin{split}
  SU(2)/U(1)_{\text{vector}}&\ =\ \bigl\{\bigl(r\,e^{i\phi},\sqrt{1-r^2}\bigr)\,
  \bigl|\,0\leq r\leq1\bigr\}\ \cong\ D^2\\[2mm]
  SU(2)/U(1)_{\text{axial}}&\ =\ \bigl\{\bigl(r,\sqrt{1-r^2}\,e^{i\phi}\bigr)\,
  \bigl|\,0\leq r\leq1\bigr\}\ \cong\ D^2\ \ .
  \end{split}
\end{equation*}
In both cases the target space is topologically given by a disc $D^2$. This
can also easily be inferred from the figures \ref{fig:SU2} and \ref{fig:SU2alt}.

Let us now proceed to the geometry of branes in this geometry. Our general 
recipe instructs us to search for embedding chains of some depth $N$ and 
then to pick automorphisms $\Omega_s$ for each of the groups in the chain. 
Here our chains will have length $N=2$, they consist of $U_1 = U(1)$ and 
$U_2 = SU(2)$ with some homomorphism $\epsilon: U_1 \rightarrow SU(2)$. On 
$U_1 = U(1)$ there exist two different automorphisms $\Omega_1$, namely the
identity $\id$ and the inversion $\gamma$. The latter sends each $\eta 
\in U(1)$ to its inverse $\gamma(\eta) = \eta^{-1}$. Automorphisms $\Omega_2$ 
of $SU(2)$ are all inner so that they are parametrized by elements of $SU(2)$.
As we discussed above, these data have to obey the two conditions (\ref{cond1},\,\ref{cond2}). 
In our situation this means that $\epsilon = \epsilon_L$ and $\Omega_2\circ
\epsilon\circ\Omega_1=\epsilon_R$. To describe solutions of the second 
condition we introduce the following conjugation on $SU(2)$, 
\begin{equation}
  \label{eq:TwistSU}
  \gamma\, \mat z_1&z_2\\-\bar{z}_2&\bar{z}_1\tam  \ = \ 
   \overline{\mat z_1&z_2\\-\bar{z}_2&\bar{z}_1\tam}
  \ =\ \mat0&1\\-1&0\tam\mat z_1&z_2\\-\bar{z}_2&\bar{z}_1\tam\mat0&-1\\1&0\tam\ \ .
\end{equation}
In the case of the vector gauging, $\epsilon_R = \epsilon_L$ and hence we look 
for pairs of $\Omega_1,\Omega_2$ such that $\Omega_2 \circ \epsilon \circ \Omega_1
= \epsilon$. This is satisfied for $(\Omega_1,\Omega_2) = (\id,\id)$ and for 
$(\Omega_1,\Omega_2) = (\gamma, \gamma)$. While the first choice gives $A$-branes, 
the second is associated with $B$-branes in the terminology of \cite{Maldacena:2001ky}. 
The analysis of axial gauging is similar and leads to the two possibilities 
$(\Omega_1,\Omega_2) = (\id,\gamma)$ and $(\Omega_1,\Omega_2) = (\gamma, \id)$.%
\smallskip 
 
To describe the D-branes in the parafermion theory we apply our general scheme 
according to which we have to consider products of twisted conjugacy classes
\begin{equation*}
  \mc{C}_{\mu}^{SU(2)}(\Omega_2)\cdot\Omega_2\circ\epsilon\bigl(\mc{C}^{U(1)}_a
   (\Omega_1)\bigr) \ =\ %
  \Bigl\{gg_\mu\Omega_2(g^{-1})\cdot\Omega_2\circ\epsilon\bigl(hh_a 
  \Omega_1(h^{-1})\bigr) \Bigr\}\ \ .
\end{equation*}
Here, $g_\mu \in SU(2)$ and $h_a \in U(1)$ are two fixed elements and $g\in 
SU(2), h \in U(1)$ are allowed to run over the whole groups. To make this more 
explicit let us restrict to the case of vector gauging. For the $A$-branes one 
can easily see that the relevant conjugacy classes have the form (with $2c=\tr\,g_\mu$)
\begin{equation*}
  \begin{split}
  \mc{C}_{\mu}^{SU(2)}(\id)&\ =\ %
  \bigl\{\bigl(\,c\pm i\sqrt{r^2-c^2},\sqrt{1-r^2}\,e^{i\phi_2}\bigr)\,\bigl|
   \ \ |c|\leq r\leq1\bigr\}\\[2mm]
  \epsilon\bigl(\mc{C}^{U(1)}_a(\id)\bigr)&\ =\ \bigl\{(e^{ia},0)\bigr\}\ \ .
  \end{split}
\end{equation*}
The $A$-branes are then parametrized by
\begin{equation*}
  \mc{C}_{\mu}^{SU(2)}\cdot\epsilon\bigl(\mc{C}^{U(1)}_a\bigr)\ =\ %
  \bigl\{\bigl((c\pm i\sqrt{r^2-c^2})e^{ia},\sqrt{1-r^2}\,e^{i(\phi_2-a)}\bigr)
  \,\bigl|\ \ |c|\leq r\leq1\bigr\}\ \ .
\end{equation*}
It is now very easy to depict these branes in the figures \ref{fig:SU2}
and \ref{fig:SU2alt}. After vector gauging we recover one-dimensional 
branes stretching between two points on the boundary of the disc. For 
$c=1$ they degenerate to a point-like object on the boundary.%
\smallskip

In the case of $B$-branes, we use the following twisted conjugacy classes 
on $SU(2)$ and $U(1)$ in our construction  
\begin{eqnarray*}
  \mc{C}_{\mu}^{SU(2)}(\gamma)&=& %
  \bigl\{\bigl(r\,e^{i\phi_1},c\pm i\sqrt{1-r^2-c^2}\bigr)\bigl|\,
  0\leq r\leq\sqrt{1-c^2}\leq1\bigr\}\\[2mm]
  \gamma \circ\epsilon\bigl(\mc{C}^{U(1)}_a(\gamma)\bigr)&=&
  \bigl\{(e^{i\xi},0)\bigl|\,0\leq\xi\leq2\pi\bigr\}\ \ .
\end{eqnarray*}
If we take the product, the branes obviously fill the whole space in
the figures \ref{fig:SU2} and \ref{fig:SU2alt} -- but only for values 
$0\leq r\leq\sqrt{1-c^2}\leq1$. After vector gauging we thus recover 
the usual $B$-branes which are two-dimensional discs centered around 
the origin of the target space disc. For $c=0$ they degenerate to a 
truly space filling brane. It is not difficult to work out the 
corresponding results for axial gauging.%

\subsection{$T^{pq}$ spaces and the conifold}

The spaces $T^{pq}$ that we are about to analyze next are simple 
generalization of the space $T^{11}$. The latter is a close relative of 
the base of the conifold  in which the RR-fluxes of the latter are replaced 
by a NSNS background field \cite{Pando-Zayas:2000he}. Our general theory 
provides a large class of boundary theories for this background, 
including branes that wrap one of the three-spheres in $T^{11}$. Related 
objects play an important role in the conifold geometry.%
\smallskip

The $T^{pq}$ spaces are defined to be quotients $SU(2)_{k_1}\times 
SU(2)_{k_2}/U(1)_k$ where the $U(1)$ subgroup acts by twisted conjugation, 
i.e.\ according to $(g_1,g_2)\mapsto\bigl(g_1\epsilon_p(h^{-1}),\epsilon_q(h)g_2
\bigr)$ where $\epsilon_p (\eta) = \epsilon (\eta^p)$ and $\epsilon$ is the usual 
embedding of $U(1)$ into $SU(2)$. We obtain this action from the choice 
\begin{equation*}
  \epsilon_L = e \times \epsilon_q \ \ \ , \ \ \ 
  \epsilon_R = \epsilon_p \times e\ \ .
\end{equation*}
If we parametrize the first factor $SU(2)$ by $(z_1,z_2)$ as before and similarly 
use $(z_1',z'_2)$ for the second factor, we realize that the action of
$\eta = \exp(i \tau) \in U(1)$ can be stated more explicitly by the formula
\begin{equation}
  \label{eq:Gauging}
  (z_1,z_2,z_1^\prime,z_2^\prime)\ \mapsto\ %
  (e^{-ip\tau}z_1,e^{ip\tau}z_2,e^{iq\tau}z_1^\prime,e^{iq\tau}z_2^\prime)\ \ .
\end{equation}
The corresponding gauged WZNW functional is free of anomalies provided that
$k=k_1p^2=k_2q^2$ (see condition \eqref{eq:NoAnomaly}). Note that the resulting coset 
still has a $SU(2)\times SU(2)$ symmetry which is realized by $(g_1,g_2)
\mapsto\bigl(h_1g_1,g_2h_2\bigr)$.%
\smallskip
  
The geometry of the coset may be deduced from the action \eqref{eq:Gauging}
by putting $z_1$ to a positive real number. This works fine except for $z_1=0$ where 
we have to gauge $z_2$ for instance. The resulting geometry is based on a 
product of a two-sphere times a three-sphere. Due to the non-trivial embeddings 
only part of the $U(1)$ has to be used for the gauging and a detailed analysis
yields 
\begin{equation*}
  SU(2)\times SU(2)/U(1)\ =\ (S^2\times S^3)/\Integer_p\ \ .%
\end{equation*}
\medskip
  
Let us now have a look for the D-branes in this geometry. According to the 
general procedure we are instructed to select a chain of groups. Here we will 
work with a chain of length $N=3$ consisting of $U_1 = U(1)$, $U_2 = U(1) \times
U(1)$ and $U_3 = SU(2)\times SU(2)$. We also pick embedding maps $\epsilon_1: U(1) 
\rightarrow U(1) \times U(1)$ defined by $\epsilon_1(\eta) = e \times \eta$ 
and $\epsilon_2: U(1) \times U(1) \rightarrow SU(2) \times SU(2)$ given 
through $ \epsilon_2 = \epsilon_p \times \epsilon_q$. Furthermore, we shall 
assume that the automorphism $\Omega_1$ is the identity map. The 
other two automorphisms $\Omega_2 = \Omega'$ and $\Omega_3 = \Omega$ 
are allowed to be non-trivial. D-branes obtained from these data wrap the 
following product of twisted conjugacy classes,
\begin{equation*}
  \mc{C}^{SU(2)\times SU(2)}(\Omega)\cdot\Omega\circ\epsilon_L\bigl(\mc{C}^{U(1)\times U(1)}
  (\Omega^\prime)\bigr)\ \ .
\end{equation*}
In writing this formula, we omitted the factor associated with a conjugacy 
class in $U_1 = U(1)$. Conjugacy classes of $U(1)$ consist of a single point 
which we can choose to be the unit element. The sets above descend to the 
coset space $T^{pq}$ if $\Omega\circ\epsilon_L\circ\Omega^\prime=\epsilon_R$ 
(note that the condition $\epsilon_2 \circ \epsilon_1 = \epsilon_L$ holds by 
construction). One solution to this condition is $\Omega = \id \times \id$ and
$\Omega' = \sigma$ the permutation of the two $U(1)$ factors.%
\smallskip

Given these gluing automorphisms, the relevant conjugacy classes in $SU(2) 
\times SU(2)$ are given by
\begin{multline*}
  \mc{C}_\mu^{SU(2)}\times\mc{C}_\nu^{SU(2)}\\\ =\ %
  \bigl\{\bigl(re^{\pm i\theta(r)},\sqrt{1-r^2}e^{i\phi_2},r^\prime 
  e^{\pm i\theta^\prime(r^\prime)},\sqrt{1-r^{\prime2}}
  e^{i\phi_2^\prime}\bigr)\, \bigl|\ |c|\leq r\leq1\ ,
  |c^\prime|\leq r^\prime\leq1\bigr\}
\end{multline*}
where the signs $\pm$ may be chosen independently and $r\cos\theta(r)=c_\mu=
\tr\,g_\mu/2$ as well as $r^\prime\cos\theta^\prime(r^\prime)=c_\nu=\tr\,g_\nu/2$.
These equations may only be solved for $r\geq|c_\mu|$ and $r^\prime\geq 
|c_\nu|$. The twisted conjugacy classes in $U(1) \times U(1)$ are of the 
form 
\begin{equation*}
  \mc{C}_a^{U(1)\times U(1)}(\Omega = \sigma)\ =\ %
  \{\bigl(e^{i(a+\xi)},e^{i(a-\xi)}\bigr)\}\ \stackrel{\epsilon_L}{\mapsto}\ 
  \{\bigl(e^{i(a+\xi)},0,e^{i(a-\xi)},0\bigr)\}\ \ .
\end{equation*}
Combining these two results we find the following expression for the product  
\begin{multline*}
  \mc{C}_\mu^{SU(2)}\times\mc{C}_\nu^{SU(2)}\cdot\epsilon\bigl(\mc{C}_a^{U(1)\times U(1)}(\Omega=\sigma)\bigr)\
\\[2mm] = \ 
  \bigl\{\bigl(re^{\pm i\theta(r)+ip(a+\xi)},\sqrt{1-r^2}e^{i(\phi_2-p(a+\xi))},
  r^\prime e^{\pm i\theta^\prime(r^\prime)+iq(a-\xi)},\sqrt{1-r^{\prime2}}
  e^{i(\phi_2^\prime-q(a-\xi))}\bigr)\bigr\}
\end{multline*}
We may use the gauge freedom to put the first entry to $r$. This is equivalent 
to setting $p\tau=\pm\theta(r)+p(a+\xi)$ in eq.\ \eqref{eq:Gauging}. The resulting
terms in the second and fourth entry may be compensated by a redefinition of
$\phi_2$ and $\phi_2^\prime$. We are thus left with
\begin{multline*}
  \mc{C}_\mu^{SU(2)}\times\mc{C}_\nu^{SU(2)}\cdot\epsilon\bigl(\mc{C}_a^{U(1)\times U(1)}(\Omega=\sigma)\bigr)/U(1)
\\[2mm] =\ %
  \bigl\{\bigl(r,\sqrt{1-r^2}e^{i\phi_2},r^\prime 
e^{\pm i\theta^\prime(r^\prime)\pm iq/p\theta(r)+2iqa},\sqrt{1-r^{\prime2}}
e^{i\phi_2^\prime}\bigr)\bigr\}\ \ .
\end{multline*}
Let us point out that after gauging we eliminated the variable $\xi$ that 
parametrized the twisted conjugacy classes in $U(1) \times U(1)$. Hence, 
these branes in $T^{pq}$ have the same dimensionality as conjugacy classes 
in $SU(2) \times SU(2)$, i.e.\ they are $0,2$ or $4-$dimensional.%
\smallskip 

We can also construct odd dimensional branes in $T^{pq}$ but this requires 
to change some of the data we have been using. We stay with the same groups 
$U_s$ and embeddings as above but choose a different collection of automorphisms. 
For the group $SU(2)\times SU(2)$ we use the non-trivial inner automorphism
$\Omega_3=(\gamma,\gamma)$ whose constituents have been defined in eq.\ 
\eqref{eq:TwistSU}. The condition \eqref{cond2} may then be fulfilled if the 
automorphism $\Omega_2$ of $U(1)\times U(1)$ is given by the exchange of group 
factors and $\Omega_1$ by the inversion $\gamma$.%
\smallskip

Under these circumstances, the twisted conjugacy classes in the group $SU(2)
\times SU(2)$ are typically four-dimensional submanifolds of the form $S^2
\times S^2$ while those of $U(1)\times U(1)$ and $U(1)$ are both one-dimensional. 
One may easily see that the product of them inside $SU(2)\times SU(2)$ is a 
submanifold of dimension $2$, $4$ or $6$. After gauging the $U(1)$ we are thus 
left with all kinds of odd-dimensional branes. When the levels are even, it 
is possible to find three-dimensional branes which fill one of the three-spheres of 
$T^{pq}$. Related objects play an important role for string theory on the 
conifold.%
  
\subsection{The big-bang big-crunch scenario}

\begin{figure}
  \centerline{\input{SL2.pstex_t}}
  \caption{\label{fig:SL2}The group manifold $SL(2,\Real)$.}
\end{figure}
  
Recently, there has been renewed interest \cite{Elitzur:2002rt} in the
Nappi-Witten background \cite{Nappi:1992kv} which describes a closed
universe between a big-bang and a big-crunch singularity. It was shown 
that the dynamics couples the closed universe to regions in space-time 
which formerly were believed to be unphysical. The full geometry is given 
by the coset $SL(2,\Real)\times SU(2)/\Real\times\Real$
where the groups in the numerator act asymmetrically on both 
factors in the denominator. Here we shall apply our general framework 
to the discussion of brane geometries in these asymmetric cosets. We 
believe that the construction of the corresponding boundary states in
these non-compact backgrounds is possible using results from 
\cite{Ponsot:2001gt,Lee:2001gh}.%
\smallskip 

Let us review the geometry of the target space first. For our purposes it
is convenient to parametrize the group manifold $SL(2,\Real)$ according to
\begin{equation}
  \label{eq:DefSL}
  \mat X_0+X_3&X_1+X_2\\X_1-X_2&X_0-X_3\tam\ \ \text{ with }\ \ 
X_0^2-X_1^2+X_2^2-X_3^2=1,\ \ X_i\in\Real\ \ .
\end{equation}
In close analogy to the case of $SU(2)$, this set can be depicted as a product 
of hyperbolas $X_1^2-X_2^2=r$ and $X_0^2-X_3^2=1+r$ in the $(X_1,X_2)$-plane and
the $(X_0,X_3)$-plane, respectively. These hyperbolas are fibered over the real 
coordinate $r$ and they degenerate in one of the two planes for $r=-1,0$. We thus 
have to distinguish the regions $r>0$, $0>r>-1$ and $-1>r$. The resulting geometry
is pictured in figure \ref{fig:SL2} as a fibre over $r\in\Real$. The parametrization 
of $SU(2)$ has already been given in section \ref{sc:Parafermions} (see figures 
\ref{fig:SU2} and \ref{fig:SU2alt}).%
\smallskip 

\begin{figure}
  \centerline{\input{SL2gauged.pstex_t}}
  \caption{\label{fig:SL2gauge1}The group manifold $SL(2,\Real)$ after gauging.}
\end{figure}
  
\begin{figure}
  \centerline{\input{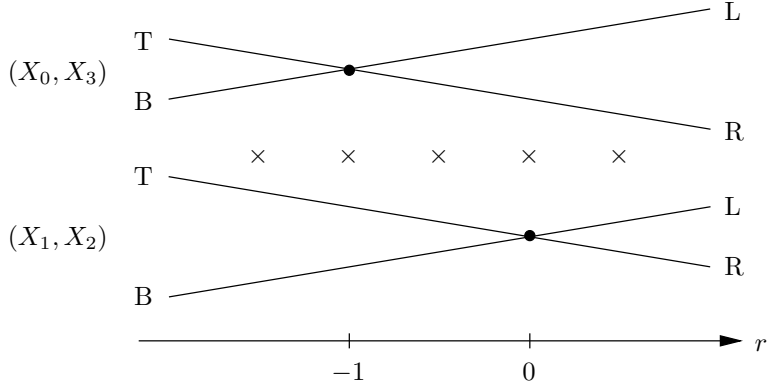}}
  \caption{\label{fig:SL2gauge2}An alternative representation of the group manifold $SL(2,\Real)$ after gauging.}
\end{figure}

In the next step we have to specify the action of the subgroup $\Real\times\Real$ 
on $SL(2,\Real)\times SU(2)$. To make contact with the general setting of section 
\ref{sc:Bulk}
let us introduce the notation $G=G_1\times G_2=SL(2,\Real)\times SU(2)$ and $H=
\Real\times\Real$. The coset we want to consider is defined by using the 
identification $g\sim\epsilon_L(h)g\epsilon_R(h^{-1})$ where the left and right 
homomorphisms of the subgroup $H$ are defined by
  \cite{Nappi:1992kv}
\begin{eqnarray*}
  \epsilon_L(\rho,\tau)
  &=&\mat e^{\rho}&0\\0&e^{-\rho}\tam\times\mat e^{i\tau}&0\\0&e^{-i\tau}\tam
  \\[2mm]
  \epsilon_R(\rho,\tau)
  &=&\mat e^{-\tau}&0\\0&e^{\tau}\tam\times\mat e^{-i\rho}&0\\0&e^{i\rho}\tam\ \ .
\end{eqnarray*}
Using these expressions it is not difficult to see that the action of $H$ leaves 
the quantities $X_0^2-X_3^2$, $X_1^2-X_2^2$, $|z_1|$ and $|z_2|$ invariant. In 
fact, these transformations correspond to boosts on the hyperbolas and rotations 
on the circles. Deviating from the analysis in \cite{Elitzur:2002rt} we will 
perform the gauge fixing completely in the $SL(2,\Real)$ part of the target 
space. As can easily be seen, the gauge transformations allow to gauge the
$SL(2,\Real)$ hyperbolas down to two disconnected points.
This procedure completely removes the gauge freedom 
except for singular points at $r=-1,0$. These points correspond to the big-bang 
and big-crunch singularities and we will not be concerned too much with details 
of the geometry at these special points. The findings of these considerations 
are illustrated in the figures \ref{fig:SL2gauge1} and \ref{fig:SL2gauge2}.%
\smallskip

\begin{figure}
  \centerline{\input{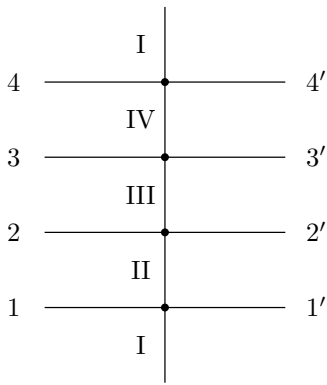}}
  \caption{\label{fig:BigBangBigCrunch}The big-bang/big-crunch scenario.}
\end{figure}

\begin{figure}
  \centerline{\input{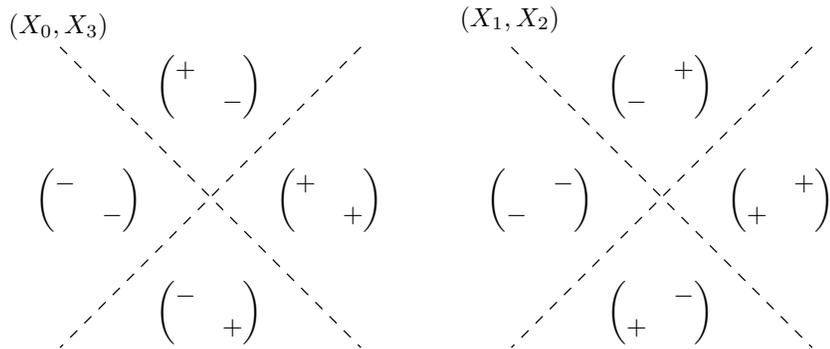}}
  \caption{\label{fig:regions}Different regions of $SL(2,\Real)$ and where they appear in our picture.
  The matrix elements indicate the sign of $X_0\pm X_3$ and $X_1\pm X_2$, respectively.}
\end{figure}

It is now only a short step to recover the results of \cite{Elitzur:2002rt}.
Let us introduce the notation $L,R,T,B$ which are shorthand for left, right,
top and bottom and specify the location of points in figure \ref{fig:SL2gauge1}. 
The regions of $SL(2,\Real)$ which appear in the fibre over $r\in\Real$ can be 
described by pairs of symbols $L,R,T,B$. A short look at figure \ref{fig:SL2gauge1}
reveals that only twelve different combinations are allowed. Working out the 
connectivity properties of these 
different regions we arrive at figure \ref{fig:BigBangBigCrunch} which has 
also been obtained in \cite{Elitzur:2002rt}. In order to simplify the comparison 
with \cite{Elitzur:2002rt} we have adopted their notation. The translation can be 
performed by means of table \ref{tb:Translation} (see also figure \ref{fig:regions}). 
From figure \ref{fig:BigBangBigCrunch} we observe that there are four closed compact 
universes I--IV which are connected at the big-bang and big-crunch singularities. At 
each instant of time they have the topology of a three-sphere $S^3$ if one takes the 
$SU(2)$ factor into account. The periodicity of time may be resolved by going to the 
infinite cover $AdS_3$ of $SL(2,\Real)$. In addition to the closed universes there 
are eight whiskers which are also connected to the singularities. Over each point 
in the whisker one has a $S^3$.%
\medskip

Let us now begin to place branes into this geometry. Once more we shall work 
with chains of length $N=2$ and the homomorphism $\epsilon = \epsilon_L: 
U_1 = \Real \times \Real \rightarrow U_2 = SL(2,\Real) \times SU(2)$. If we 
define an automorphism $\Omega$ of $\Real \times \Real$ by $\Omega(\tau, \rho) 
= (-\rho,-\tau)$ then our condition \eqref{cond2} is satisfied whenever 
\begin{equation*}
  \Omega_2 \circ \epsilon \circ \Omega_1 = \epsilon \circ \Omega \ \ .
\end{equation*}
D-branes in our background should be localized along the following product of 
twisted conjugacy classes,
\begin{equation}
  \label{eq:BCBrane}
  \Bigl[\mc{C}_\mu^{SL(2,\Real)}(\omega_1)\times\mc{C}_\nu^{SU(2)}(\omega'_1)\Bigr]
  \cdot(\omega_1\times\omega'_1)\circ\epsilon\bigl(\mc{C}_{a}^{\Real\times\Real}
  (\Omega_2)\bigr)\ \ ,
\end{equation}
before projecting to the coset. Here, we split $\Omega_1  = \omega_1 \times \omega'_1$ 
into the product of automorphisms for $SL(2,\Real)$ and $SU(2)$, respectively. There 
are several choices of automorphisms $\Omega_2,\omega_1,\omega'_1$ which satisfy our 
condition and we will discuss all of them in the following.%
\smallskip 

Let us start with the discussion of the twisted conjugacy class
$\mc{C}_{a}^{\Real\times\Real}(\Omega_2)$. The most general automorphism
of the additive group $\Real\times\Real$ is implemented by a non-singular
$2\times 2$-matrix. In our situation, however, not all choices are allowed.
The only choices which have the chance to be consistent with condition
\eqref{cond2} are $\Omega_2(\rho,\tau)=(\eta\tau,\xi\rho)$ where
$\eta,\xi=\pm 1$. The resulting geometry is given by
\begin{equation}
  \label{eq:TCC}
  \mc{C}_{a}^{\Real\times\Real}(\Omega_2)
  \ =\ \left\{\begin{array}{cl}
         \Real\times\Real&,\,\text{for }\xi=-\eta\\[2mm]
         \bigl\{(f_1+\lambda,f_2-\eta\lambda)\,\bigl|\,\lambda\in\Real\bigr\}
          &,\,\text{for }\xi=\eta\ \ .
       \end{array}\right.
\end{equation}
The embedding of these sets into $SL(2,\Real)\times SU(2)$ via the map
$(\omega_1\times\omega'_1)\circ\epsilon$ leads to the same result in both
cases after gauge fixing.%
\smallskip

When investigating the geometry of the D-branes \eqref{eq:BCBrane} in the
big-bang big-crunch target space it is convenient to focus on the $SL(2,\Real)$ 
part as all interesting features arise from this factor. We thus only have to 
distinguish two different cases, corresponding to the two types of twisted 
conjugacy classes of $SL(2,\Real)$. As most of the group $SL(2,\Real)$ will 
be gauged away, it suffices to address the following two questions:
\begin{enumerate}
\item Which ranges of $r$ are covered by the twisted conjugacy classes?
\item Does the twisted conjugacy class extend along one or even both branches 
  of the hyperbolas, i.e.\ does the D-brane cover one or two points for fixed 
  value of $r$ after gauging?
\end{enumerate}
\smallskip 
  
The twisted conjugacy classes of $SL(2,\Real)$ are easily described.
For untwisted conjugacy classes one 
has two types. There are two point-like conjugacy classes which correspond 
to the center of $SL(2,\Real)$ while all others are two-dimensional. The 
exact shape has been worked out in \cite{Stanciu:1999nx,Bachas:2000fr} but 
we will not need these details. The point-like branes are specified by 
$X_0=\pm1$ and $X_1=X_2=X_3=0$, i.e.\ they are localized at $r=0$. After 
gauging they sit at the singularities between the closed universes I--II 
and III--IV, respectively. The two-dimensional conjugacy classes are of 
the form $X_0=C=\text{const.}$ with arbitrary values of the remaining 
coordinates. According to the constraint \eqref{eq:DefSL} we obtain
$r=C^2-1-X_3^2\leq C^2-1$. This means that the conjugacy class after gauging
covers at least all four whiskers $2,2^\prime,4,4^\prime$. For $C\neq0$ the 
conjugacy class grows into two of the four closed universes starting from 
the singularity which joins them. Depending on the sign of $C$ these are the 
regions I--II (for $C>0$) and III--IV (for $C<0$). If $|C|$ reaches the value 
$1$ (from below) the conjugacy class stretches completely through both of the 
closed universes. Increasing $|C|$ further, the conjugacy classes start to 
reach into two of the remaining whiskers -- $1,1^\prime$ for $C>1$ and 
$3,3^\prime$ for $C<-1$. Note, that the multiplication with the twisted 
conjugacy class of $\Real\times\Real$ has no influence on the possible 
values of $r$ as it simply corresponds to some boost on the hyperbolas 
which will be gauged away in any case.%
\smallskip 
  
\begin{table}
  \centerline{\begin{tabular}{cccccccccccc}
  (R,B) & (R,T) & (L,T) & (L,B) & (R,R) & (R,L) & (B,T) & (T,T) & (L,L) & (L,R) & (T,B) & (B,B)\\\hline
   I & II & III & IV & $1$ & $1^\prime$ & $2$ & $2^\prime$ & $3$ & $3^\prime$ & $4$ & $4^\prime$
  \end{tabular}}
  \caption{\label{tb:Translation}Translation table for the twelve different regions.}
\end{table}

\begin{figure}
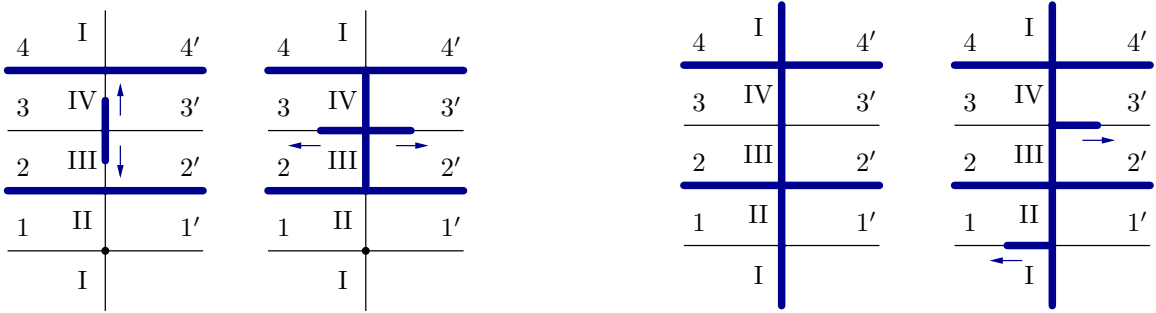

  \centerline{\input{BCBrane1.pstex_t}\quad\input{BCBrane2.pstex_t}\qquad\qquad\qquad\input{BCBrane3.pstex_t}
  \quad\input{BCBrane4.pstex_t}}
  \caption{\label{fig:BCbranes}D-branes in the big-bang big-crunch scenario. The branes
  on the left hand side have been constructed with an ordinary conjugacy class of $SL(2,\Real)$
  while for the right ones a twisted conjugacy class was employed.}
\end{figure}

The twisted conjugacy classes arise from the automorphism which
reverses the sign of $X_2$ and $X_3$. It may be described by conjugation
with the element $M=\left(\begin{smallmatrix}0&1\\1&0\end{smallmatrix}
\right)$. The corresponding twisted conjugacy classes are given by
$\tr(Mg)=2X_1=2C=\text{const.}$ According to the constraint \eqref{eq:DefSL} 
we obtain $r=C^2-X_2^2\leq C^2$. The discussion is similar as in the untwisted 
case. For all values of $C$ the twisted conjugacy classes pass through all four 
closed universes I--IV and the four whiskers $2,2^\prime,4,4^\prime$. For
$C\neq0$ the conjugacy classes also cover part of the whiskers $1,3^\prime$
($C>0$) or $1^\prime,3$ ($C<0$). The results of the last two paragraphs are
illustrated in figure \ref{fig:BCbranes}.%
\smallskip 
  
So far we have only considered the $SL(2,\Real)$ part of the target space.
To obtain the complete picture we have also to take the $SU(2)$ part into
account as well as the product with the twisted conjugacy class $\mc{C}_{a}
^{\Real\times\Real}(\omega)$. We already argued that the latter has no effect 
on the $SL(2,\Real)$ part as it does not affect the value of $r$ and may thus 
be gauged away. This statement also implies that the resulting D-branes 
factorize (in the same sense as the gauge fixing factorized). If we try to 
solve condition \eqref{cond2} with $\omega_1=\id$, i.e.\ if we want 
to take the ordinary conjugacy classes in the $SL(2,\Real)$ part, we have to 
use an automorphism $\Omega_2$ of $\Real\times\Real$ with $\eta=1$. Depending 
on the choice of $\xi$ we are still able to obtain both expressions for 
twisted conjugacy classes that appear in eq.\ \eqref{eq:TCC}. The same 
statement holds true for $\eta=-1$, i.e.\ for the case of a twisted conjugacy 
class in the $SL(2,\Real)$ part.%
\smallskip 

It is now very simple to describe the geometry of the D-branes in the $SU(2)$
part. We simply have to multiply the (shifted) conjugacy class of $SU(2)$ with 
elements of the form $\text{diag}(e^{i\lambda},e^{-i\lambda})$ for all values 
of $\lambda$. As was observed in \cite{Maldacena:2001ky,Sarkissian:2002ie} and 
more in the spirit of our approach in \cite{Quella:2002ns}, this corresponds to 
a smearing of the original conjugacy class.%
\smallskip 
  
Let us conclude with a short summary of our results. All essential
information about the target space and about its D-branes are contained
in the figure \ref{fig:BCbranes}. While the D-branes cover the high-lightened
regions in the $SL(2,\Real)$ part, we also have a three-sphere over each of these 
points which is partly covered by the D-brane. The geometry of the latter is 
either given by a circle around some equator or by a smeared two-sphere which 
covers a three-dimensional subset of $S^3$.%

\section{Conclusions}

In this work we presented a comprehensive description of asymmetric coset models $G/H$ 
for which the action of $H$ on $G$ is not necessarily given by the adjoint. A bulk
partition 
function was proposed based on a semi-classical analysis in the large volume limit 
and the modular invariance of this partition function was shown to be equivalent to 
the anomaly cancellation that is known from the Lagrangian description of the coset.%
\smallskip

We then provided a general prescription of constructing branes in asymmetric cosets.
Due to the heterotic nature of the models, one is naturally lead to boundary
states which break part of the symmetry of the bulk theory. The geometry of
the branes may be deduced from those of symmetry breaking branes on group
manifolds \cite{Quella:2002ns}. Branes which possess a symmetry compatible
with the gauge action were argued to descend naturally to the coset.%
\smallskip

Our general findings have been used to construct D-branes in the cosmological
Nappi-Witten background (big-bang big-crunch space-time) and in the base $S^2\times S^3$ 
of the conifold. Among the branes in the big-bang big-crunch space-time there are examples 
which cross the singularities and run through all the universes. In the base of the 
conifold we found branes of all dimensions. For even values of the level one may construct 
branes which fill one of the three-spheres.%
\smallskip

Before we conclude let us mention a few open problems which remain to be solved.
Our algebraic construction of branes in asymmetric cosets was designed for the 
case that left and right action of the gauge group can be related by automorphisms 
in a chain of intermediate groups (asymmetric cosets of generalized automorphism 
type). If this condition is not fulfilled, one could be tempted to follow 
\cite{Walton:2002db} (see also \cite{Quella:2002ns}) and to propose generalized 
conjugacy classes for the geometry of the corresponding branes. While this 
procedure works fine on the Lagrangian level, we have not been able to implement 
it in an algebraic approach.%
\smallskip

Another issue is the discussion of the stability and the dynamics of
our branes. In the geometric regime the question of stability should be
accessible from a Born-Infeld analysis \cite{Maldacena:2001ky}. One may even
hope to go one step beyond such an analysis and to construct the non-commutative 
gauge theory which governs the dynamics of branes in asymmetric cosets 
\cite{Fredenhagen:2001kw}. It may also well be that the results of 
\cite{Fredenhagen:2002qn} generalize to asymmetric cosets.%

\subsubsection*{Acknowledgements}

We would like to thank B.\ Acharya, S.\ Fredenhagen, A.\ Giveon, N.\ Kim
and I.\ Runkel for remarks and discussions. One of us (T.Q.) is also grateful
to M.\ Tierz and the people at the SPhT for their kind hospitality.
This work was initiated in discussions with B.\ Acharya while one of the
authors (V.S.) was visiting the string group at Rutgers University.
The research of T.Q. was financially supported by the Studienstiftung des
deutschen Volkes.%

\begin{appendix}
\section{\label{ap:GMM}The GMM cosets revisited}

  Models of the type which have been discussed in section \ref{sc:ExamplesTPQ} first appeared
  in \cite{Guadagnini:1987ty,Guadagnini:1987qc}. These authors presented
  a Lagrangian formulation of these theories and considered the associated
  current algebra. In our opinion their discussion of the algebraic
  properties is not completely accurate. In particular they argued that 
  the energy momentum tensor is {\em not} obtained by the
  standard affine Sugawara \cite{Sugawara:1968rw,Bardakci:1971nb,Halpern:1971ay}
  and coset constructions \cite{Halpern:1971ay,Goddard:1985vk} which seems to be
  incorrect. We take this as an opportunity to review the Lagrangian description
  and to clarify some statements.%
\smallskip

  The gauged WZNW functional \eqref{eq:GaugedWZNW} is quadratic in the gauge
  fields. It may thus be simplified -- in principle -- by integrating out the
  gauge fields. The resulting expressions will, however, remain quite formal
  in the general case (see however \cite{Bardakci:1988ee,Nahm:1988sn,Gawedzki:1988hq,
  Gawedzki:1989nj,Karabali:1989au,Karabali:1990dk,Tseytlin:1994my}).
  The reason for these difficulties is the third term
  in the interaction functional \eqref{eq:Interaction} which does not only
  contain the gauge fields $A$ and $\bar{A}$ but also the group element $g$. For the
  Gaussian path integral to be performed one would need to diagonalize the
  quadratic form matrix which depends explicitly on $g$.%
\smallskip

  For our particular choice of embeddings the corresponding term vanishes and the path
  integral may easily be evaluated. The interaction functional
  \eqref{eq:Interaction} reduces to
\begin{align*}
  S_{\inta}^{G_1\times G_2/H}(g_1,g_2,A,\bar{A}|k,\epsilon_{L/R})
  &\ =\ \frac{k_1}{4\pi}\int_\Sigma\!d^2\!z
       \ \Tr_1\bigl\{-2\epsilon_1(A)g_1^{-1}\bartial g_1
                     -\epsilon_1(\bar{A})\epsilon_1(A)\bigr\}\\
   &+\frac{k_2}{4\pi}\int_\Sigma\!d^2\!z
       \ \Tr_2\bigl\{2\epsilon_2(\bar{A})\partial g_2g_2^{-1}
                     -\epsilon_2(\bar{A})\epsilon_2(A)\bigr\}\ \ .
\end{align*}
  It is fairly simple to read off the quadratic form matrix from this expression
  and integrate out the gauge fields in full generality. We only have to be
  a bit careful about our notations. We may decompose the $\h$-valued gauge
  fields $A$ and $\bar{A}$ according to $A=A_\alpha T^\alpha$ and
  $\bar{A}=\bar{A}_\alpha T^\alpha$. The {\em abstract} Lie algebra generators
  satisfy the commutation relations $[T^\alpha,T^\beta]=\ii{f^{\alpha\beta}}_\gamma T^\gamma$.
  Indices are raised and lowered using the Killing form\footnote{We remind the reader
  that $\Tr$ is a normalized trace and that we work in the conventions of \cite{Fuchs:1995}.}
\begin{equation*}
  2\,\kappa^{\alpha\beta}\ =\ \Tr\bigl\{T^\alpha T^\beta\bigr\}
\end{equation*}
  and its inverse.
  We may choose generators $T^i\in\{\epsilon_1(T^\alpha),T^I\}$ of $\g_1$ and
  generators $T^a\in\{\epsilon_2(T^\alpha),T^A\}$ of $\g_2$. These satisfy
  $[T^i,T^j]=\ii{f^{ij}}_k T^k$ and $[T^a,T^b]=\ii{f^{ab}}_c T^c$. If all three
  indices take values in the subalgebra $\h$, the structure constants by
  construction just reduce to the structure constants of $\h$ in the given
  basis. This is only true as long as the index structure is as indicated
  because one would have to use different Killing forms to lower the indices.
  From \eqref{eq:EmbIndex} it follows that they satisfy
\begin{equation*}
  \kappa^{\alpha\beta}
  \ =\ \kappa_1^{\alpha\beta}/x_1
  \ =\ \kappa_2^{\alpha\beta}/x_2\ \ .
\end{equation*}
  We see the embedding indices $x_i$ entering this expression.%
\smallskip

  The last relations imply
\begin{align*}
  \Tr_1\bigl\{\epsilon_1(\bar{A})\,\epsilon_1(A)\bigr\}
  &\ =\ 2\,x_1\,\bar{A}_\alpha A^\alpha&
  \Tr_2\bigl\{\epsilon_2(\bar{A})\,\epsilon_2(A)\bigr\}
  &\ =\ 2\,x_2\,\bar{A}_\alpha A^\alpha\ \ .
\end{align*}
  The formula
\begin{equation*}
  \int d^n\!y \: e^{-\frac{1}{2}y^T\mathbb{A}y+b^Ty}
  \ =\ \frac{(2\pi)^{n/2}}{\sqrt{\det\mathbb{A}}} \: e^{\frac{1}{2}\,b^T\mathbb{A}^{-1}b}\ \ .
\end{equation*}
  for the Gaussian path integral may thus be applied with
\begin{equation*}
  y\ =\ \mat A^\alpha\\\bar{A}^\beta\tam\qquad\mathbb{A}\ =\ \mat0&\frac{x_1k_1}{\pi}\,\kappa_{\alpha\beta}\\\frac{x_2k_2}{\pi}\,\kappa_{\alpha\beta}&0\tam\qquad
  b\ =\ \mat-\frac{k_1}{2\pi}\,\Tr_1\bigl\{\epsilon_1(T^\alpha)g_1^{-1}\bartial g_1\bigr\}\\\frac{k_2}{2\pi}\,\Tr_2\bigl\{\epsilon_2(T^\alpha)\partial g_2g_2^{-1}\bigr\}\tam\ \ .
\end{equation*}
  The matrix $\mathbb{A}$ is symmetric as by assumption $k=x_1k_1=x_2k_2$.
  It may easily be inverted. After performing the Gaussian path
  integral the interaction term reads
\begin{equation}
  \label{eq:HomogenousAction}
  S_{\inta}^{G_1\times G_2/H}(g_1,g_2|\,k,\epsilon_{L/R})
  \ =\ -\frac{\kappa_{\alpha\beta}\,k_1k_2}{4\pi\,k}\int_\Sigma\!d^2\!z\:
       \Tr_1\bigl\{\epsilon_1(T^\alpha)\,g_1^{-1}\bartial g_1\bigr\}\,
       \Tr_2\bigl\{\epsilon_2(T^\beta)\,\partial g_2g_2^{-1}\bigr\}
\end{equation}
  This is exactly the action functional which was constructed in
  \cite{Guadagnini:1987ty,Guadagnini:1987qc}.%
\smallskip
  
  The action functional \eqref{eq:HomogenousAction} possesses a number of
  very interesting and useful symmetries. By construction it is invariant
  under the infinitesimal gauge transformations
  $(g_1,g_2)\mapsto\bigl(g_1(1-\ii \,\epsilon_1(\Omega)),(1+\ii \,\epsilon_2(\Omega))g_2\bigr)$
  with $\Omega=\Omega(z,\bar{z})\in\h$. In addition, the model admits
  the symmetry $G_1^{L}(z)\times G_2^{R}(\bar{z})$, i.e.\ it is invariant
  under $(g_1,g_2)\mapsto\bigl(g_1^\prime(z)g_1,g_2g_2^{\prime-1}(\bar{z})\bigr)$.
  The last symmetry is generated by the currents
\begin{eqnarray*}
  J(z)&=&J_1-\frac{k_2\,\kappa_{\alpha\beta}}{2\,x_1}\ %
        g_1\epsilon_1(T^\alpha)g_1^{-1}\ %
        \Tr_2\bigl\{\epsilon_2(T^\beta)\,\partial g_2g_2^{-1}\bigr\}\\
  \bar{J}(\bar{z})&=&\bar{J}_2+\frac{k_1\,\kappa_{\alpha\beta}}{2\,x_2}\ %
        g_2^{-1}\epsilon_2(T^\beta)g_2\ %
        \Tr_1\bigl\{\epsilon_1(T^\alpha)\,g_1^{-1}\bartial g_1\bigr\}\ \ .
\end{eqnarray*}
  They satisfy $\bartial J(z)=\partial \bar{J}(\bar{z})=0$
  by the equations of motion. During the derivation we used the
  relation $x_1k_1=x_2k_2$. Note that $J$ takes values in the Lie algebra
  $\g_1$, while $\bar{J}$ is from $\g_2$. This means that the index structure
  is $J^i$, $\bar{J}^a$ which makes explicit the heterotic nature of our coset.
  Both currents are gauge invariant.%
\smallskip

  In the algebraic description of our asymmetric coset model we already assumed some
  properties which would have been expected from a straightforward generalization
  of the GKO construction. We are now able to justify this procedure more
  rigorously by working out the energy momentum tensor and the commutation
  relations of the currents. Let us start with the latter.
  It is convenient to introduce the fields
\begin{align*}
  J_1\ &=\ -k_1 \, \partial g_1g_1^{-1}&
  \bar{J}_1\ &=\ k_1 \, g_1^{-1}\bartial g_1\\ 
  J_2\ &=\ -k_2 \, \partial g_2g_2^{-1}&
  \bar{J}_2\ &=\ k_2 \, g_2^{-1}\bartial g_2
\end{align*}
  which correspond to the (former) $G_1$ and $G_2$ currents, respectively.
  In terms of these quantities one obtains
\begin{eqnarray*}
  J(z)&=&J_1+\frac{\kappa_{\alpha\beta}}{2\,x_1}\ %
        g_1\epsilon_1(T^\alpha)g_1^{-1}\ %
        \Tr_2\bigl\{\epsilon_2(T^\beta)J_2\bigr\}\\
  \bar{J}(\bar{z})&=&\bar{J}_2+\frac{\kappa_{\alpha\beta}}{2\,x_2}\ %
        g_2^{-1}\epsilon_2(T^\beta)g_2\ %
        \Tr_1\bigl\{\epsilon_1(T^\alpha)\bar{J}_1\bigr\}\ \ .
\end{eqnarray*}
  The symmetry $G_1^{L}(z)\times G_2^{R}(\bar{z})$ implies the Ward
  identities \cite[(15.40)]{FrancescoCFT}
\begin{eqnarray*}
  \delta_L^{(1)}\langle X(w,\bar{w})\rangle
  &=&-\oint\frac{dz}{2\pi i}\ \Omega_i\langle J^i(z)X(w,\bar{w})\rangle\\
  \delta_R^{(2)}\langle X(w,\bar{w})\rangle
  &=&\oint\frac{d\bar{z}}{2\pi i}\ \Omega_a\langle\bar{J}^a(\bar{z})X(w,\bar{w})\rangle\ \ ,
\end{eqnarray*}
  which are related to the transformations
  $\delta_L^{(1)}g_1=i\,\Omega_i T^i\,g_1$ and
  $\delta_R^{(2)}g_2=-i\,g_2\,\Omega_a T^a$.
  From the previous equations we may derive the non-trivial OPEs
\begin{eqnarray*}
  J^i(z)J^j(w)
  &=&\frac{i{f^{ij}}_k}{z-w}\ J^k(w)+\frac{k_1\,\kappa_1^{ij}}{(z-w)^2}\\
  J^i(z)g_1(w,\bar{w})
  &=&-\frac{\ T^ig_1(w,\bar{w})}{z-w}\\
    J^i(z)J_1^j(w,\bar{w})
  &=&\frac{i{f^{ij}}_k}{z-w}\ J_1^k(w,\bar{w})+\frac{k_1\,\kappa_1^{ij}}{(z-w)^2}\ \ .
\end{eqnarray*}
  and
\begin{eqnarray*}
  \bar{J}^a(\bar{z})\bar{J}^b(\bar{w})
  &=&\frac{i{f^{ab}}_c}{\bar{z}-\bar{w}}\ \bar{J}^c(\bar{w})+\frac{k_2\,\kappa_2^{ab}}{(\bar{z}-\bar{w})^2}\\
  \bar{J}^a(\bar{z})g_2(w,\bar{w})
  &=&\frac{g_2(w,\bar{w})T^a}{\bar{z}-\bar{w}}\\
  \bar{J}^a(\bar{z})\bar{J}_2^b(w,\bar{w})
  &=&\frac{i{f^{ab}}_c}{\bar{z}-\bar{w}}\ \bar{J}_2^c(w,\bar{w})+\frac{k_2\,\kappa_2^{ab}}{(\bar{z}-\bar{w})^2}\ \ .
\end{eqnarray*}
  All the other OPEs vanish:
\begin{eqnarray*}
  J^i(z)\bar{J}^a(\bar{w})
  &=&J^i(z)g_2(w,\bar{w})
  =J^i(z)\bar{J}_1^j(w,\bar{w})
  =J^i(z)J_2^a(w,\bar{w})
  =J^i(z)\bar{J}_2^a(w,\bar{w})=0\\
  \bar{J}^a(\bar{z})g_1(w,\bar{w})
  &=&\bar{J}^a(\bar{z})J_1^i(w,\bar{w})
  =\bar{J}^a(\bar{z})\bar{J}_1^i(w,\bar{w})
  =\bar{J}^a(\bar{z})J_2^b(w,\bar{w})=0\ \ .
\end{eqnarray*}
  Let us emphasize the asymmetry in the OPEs which already showed up in the
  algebraic construction.%
\smallskip

  Now, that the current symmetry is under control we can focus our attention
  to the conformal symmetry, i.e.\ to the energy momentum tensor. Our treatment
  will reveal the central charge to be given by a combination of affine Sugawara and
  coset construction. Both left and right moving central charge agree. Due to the
  structure of the action functional for the asymmetric coset
  the {\em classical} chiral energy momentum tensors are given by
\begin{eqnarray*}
  T\ =\ T_1+T_2+T_{\inta}\hspace{1.5cm}\text{and}\hspace{1.5cm}
  \bar{T}\ =\ \bar{T}_1+\bar{T}_2+\bar{T}_{\inta}\ \ .
\end{eqnarray*}
  The first two summands are the standard WZNW energy momentum tensors
\begin{align*}
  T_1\ &=\ \frac{1}{4\,k_1}\ \Tr_1\bigl\{J_1J_1\bigr\}&
  \bar{T}_1\ &=\ \frac{1}{4\,k_1}\ \Tr_1\bigl\{\bar{J}_1\bar{J}_1\bigr\}\\
  T_2\ &=\ \frac{1}{4\,k_2}\ \Tr_2\bigl\{J_2J_2\bigr\}&
  \bar{T}_2\ &=\ \frac{1}{4\,k_2}\ \Tr_2\bigl\{\bar{J}_2\bar{J}_2\bigr\}\ \ .
\end{align*}
  The extra summands are given by
\begin{eqnarray*}
  T_{\inta}
  &=&\frac{k_1\ \kappa_{\alpha\beta}}{4\,x_2}\ %
   \Tr_1\bigl\{\epsilon_1(T^\alpha)\,g_1^{-1}\partial g_1\bigr\}\ %
   \Tr_2\bigl\{\epsilon_2(T^\beta)\,\partial g_2 g_2^{-1}\bigr\}\\
  \bar{T}_{\inta}
  &=&\frac{k_1\ \kappa_{\alpha\beta}}{4\,x_2}\ %
   \Tr_1\bigl\{\epsilon_1(T^\alpha)\,g_1^{-1}\bartial g_1\bigr\}\ %
   \Tr_2\bigl\{\epsilon_2(T^\beta)\,\bartial g_2 g_2^{-1}\bigr\}\ \ .
\end{eqnarray*}
  It is very instructive to evaluate the expressions $\Tr_{R_1}JJ$
  and $\Tr_{R_2}\bar{J}\bar{J}$. One is then naturally lead to
\begin{eqnarray*}
  T&=&\frac{1}{2k_1}\ J_iJ^i+\frac{1}{2k_2}\ (J_2)_a(J_2)^a-\frac{1}{2x_2k_2}\ (J_2)_\alpha(J_2)^\alpha
   \ =\ T_{k_1}^{G_1}+T_{k_2}^{G_2}-T_{x_2k_2}^H\\
  \bar{T}&=&\frac{1}{2k_2}\ \bar{J}_a\bar{J}^a+\frac{1}{2k_1}\ (J_1)_i(J_1)^i-\frac{1}{2x_1k_1}\ (J_1)_\alpha(J_1)^\alpha
   \ =\ \bar{T}_{k_1}^{G_1}+\bar{T}_{k_2}^{G_2}-\bar{T}_{x_1k_1}^H\ \ .
\end{eqnarray*}
  The additional factors $x_1$ and $x_2$ arise due to the usage of the natural
  Killing form for $\h$-quantities. After quantizing the theory the levels get
  shifted by the respective dual Coxeter numbers. Let us emphasize the following
  remarkable fact: Due to the condition $x_1k_1=x_2k_2$
  left and right moving Virasoro algebra possess the same central charge.
  This result also has been noted in \cite{Guadagnini:1987ty,Guadagnini:1987qc}
  but the algebraic reasons remained unclear. In particular  in the last
  reference due to usage of
  intransparent notation it was not realized that the energy momentum tensor
  is actually defined by the standard affine Sugawara construction combined with
  the coset construction.%

\end{appendix}
  
\providecommand{\href}[2]{#2}\begingroup\raggedright\endgroup


\end{document}